\documentclass[12pt]{article}
\usepackage{amssymb,amsmath,amsfonts,authblk}
\usepackage{graphicx}
\usepackage{color}
\usepackage{multicol}
\usepackage{float}
\usepackage[unicode=true,bookmarks=false,breaklinks=false,pdfborder={0 0 1},backref=page,colorlinks=true]{hyperref}

\usepackage{subcaption}

\newcommand{\s}{{\rm S}}
\renewcommand{\b}{{\rm B}}

\newcommand{\N}{N_{\rm tot}}

\newcommand{\h}{{\mathcal H}}
\newcommand{\rx}{{\mathbb R}}
\newcommand{\tr}{{\rm tr}}
\newcommand{\e}{{\mathcal E}}

\newcommand{\hW}{h_{W}}

\newcommand{\bbbone}{{\mathchoice {\rm 1\mskip-4mu l} {\rm 1\mskip-4mu l}{\rm 1\mskip-4.5mu l} {\rm 1\mskip-5mu l}}}

\newcounter{resultcounter}[section]

\newtheorem{definition}[resultcounter]{Definition}

\def\bed{\begin{definition}}
	\def\eed{\end{definition}}

\begin{document}
	\title{
Entropy and entanglement in a bipartite\\ quasi-Hermitian 
		system and its Hermitian counterparts}
 \author{Abed Alsalam Abu Moise \quad   Graham Cox \quad Marco Merkli\footnote{aaabumoise@mun.ca, gcox@mun.ca, merkli@mun.ca}}
 \affil{ Department of Mathematics and Statistics\\
 	Memorial University of Newfoundland\\
 St. John's, NL A1C 5S7 \\
 Canada}
\date{}
\maketitle

\begin{abstract}
We consider a quantum  oscillator coupled to a bath of $N$ other oscillators. The total system evolves with a quasi-Hermitian Hamiltonian. Associated to it is a family of Hermitian systems, 
parameterized by a unitary map $W$. Our main goal is to find the influence of $W$ on the entropy and the entanglement in the Hermitian systems. We calculate explicitly the reduced density matrix of the single oscillator for all Hermitian systems and show that, regardless of $W$, their von Neumann entropy oscillates with a common period which is twice that of the non-Hermitian system. We show that generically, the oscillator and the bath are entangled for almost all times. While the amount of entanglement depends on the choice of $W$, it is independent of $W$ when averaged over a period. These results describe some universality in the physical properties of all Hermitian systems associated to a given non-Hermitian one. 
\end{abstract}

\section{Motivation \& outline of main results}

In recent years there has been much interest in extensions of quantum mechanics that allow for non-Hermitian Hamiltonians. In $PT$-symmetric quantum theory, for instance, non-Hermitian Hamiltonians with purely real spectrum are often encountered. 
In many cases those Hamiltonians are quasi-Hermitian
\cite{BeBo1,BeBo2,Bender2003, Dieu,KS,Mos2002I, Mos2002II, Mos2002III,Mos2010,SGH}. For these one can produce associated Hermitian Hamiltonians which can then be studied using `standard' methods of quantum theory.

The assignment of a Hermitian to a non-Hermitian system is called the {\em Dyson map} \cite{FF17.2, Dyson56-1, Dyson56-2}. The Dyson map is determined up to a unitary. To a given non-Hermitian Hamiltonian $H$ are assigned Hermitian Hamiltonians $h$ which are all unitarily equivalent. In this sense, physical properties of the original $H$ are uniquely encoded by any associated $h$. However, if the system has a subsystem structure, then the unitaries generally reshuffle local degrees of freedom and the choice of the unitary plays a crucial physical role. This is in particular the case for open systems, where system and bath degrees of freedom are reshuffled. One should then ask what relevant information about the original non-Hermitian system is encoded in `the' associated Hermitian one.  In fact, if one allows the Dyson map to be time-dependent, as is often the approach taken in the literature  \cite{FF17, FF17.2, FM2016}, then the above mentioned non-uniqueness is compounded by additional freedom. It turns out that given any quasi-Hermitian system $H$, one can chose a suitable time-dependent Dyson map so that the associated Hermitian system has the trivial Hamiltonian $h=0$. We explain this in Appendix \ref{sec:dysonmap}.

The ambiguity in the associated Hermitian systems is the subject of the present paper. In the recent literature on $PT$-symmetric quantum theory it is suggested that instead of studying the entropy of an open quasi-Hermitian systems, one may study the entropy of the corresponding associated Hermitian system \cite{FF19,FrithPHD2020,Frith2020,Cen2020}. Concretely, in \cite{FF19}, the authors examine the entropy of an oscillator coupled to an `environment'  of $N$ other oscillators via a $PT$-symmetric Hamiltonian. As a proxy for the von Neumann entropy of the single oscillator in the non-Hermitian system, they study the corresponding quantity for {\em one particular choice} of an associated Hermitian system, constructed by a time-dependent Dyson map. The authors find a different qualitative behaviour of the entropy of the Hermitian system according to the $PT$-symmetry phases of the non-Hermitian system. This interesting result is found for  one particular choice of the associated Hermitian system -- the choice may be regarded as a good one as it discerns the different symmetry phases by the qualitative form of the dynamics of the entropy. For a different choice however, the entropy dynamics would look entirely different (and in particular, it would be time-independent upon chosing $h=0$, as explained above). 

In the current work we investigate {\em all} associated systems in the regime of unbroken $PT$-symmetry -- in which the non-Hermitian Hamiltonian is actually quasi-Hermitian. We vary over all time-independent Dyson maps (excluding time dependent ones for the reason mentioned above). 
\medskip

A Hamiltonian $H$ is said to be \emph{quasi-Hermitian} if 
$$
H^\dag=\eta H\eta^{-1}
$$ 
for some bounded, positive operator $\eta>0$, called a \emph{metric operator}. One obtains a Hermitian Hamiltonian by 
$$
h = SHS^{-1},
$$
where $S$ is any operator with 
$$
S^\dag S = \eta.
$$ 
The assignment $H\mapsto h$ is called {\em the Dyson map} \cite{Dyson56-1, Dyson56-2}, even though to a single $H$ one can associate  different $h$ for the following two reasons. 
\begin{itemize}
\item[--] There are different metric operators $\eta$  for the same quasi-Hermitian $H$. 

\item[--] Once a metric $\eta$ is fixed, the general solution of the equation $S^\dag S = \eta$ is $S=W\sqrt\eta$, where $\sqrt\eta$ is the unique positive operator that squares to $\eta$ and $W$ is an arbitrary unitary $W$.
\end{itemize}
The metric $\eta$ is uniquely determined if one fixes a sufficiently rich (irreducible) set of quasi-Hermitian operators (including $H$) to be {\em observables}; see \cite{SGH} and Section \ref{sec:qhs}. Once this choice is made, all associated Hermitian systems are parameterized by the unitary $W$. 
Different choices of $W$ give different Hermitian $h$, which are unitarily equivalent. However, generally $W$ reshuffles the degrees of freedom and modifies local physical properties, such as entanglement between subsystems. 

In this work, we consider a concrete quasi-Hermitian system with Hamiltonian $H$, describing the interaction of a single oscillator with a bath of $N$ other  oscillators. This is an open quantum system of the `system--bath' ($\s\b$) type, described by a bipartite Hilbert space $\h_\s\otimes\h_\b$. We take the metric $\eta$ to be general but fixed, and then analyze all the associated Hermitian systems resulting from varying over all choices of $W$. The dynamics of the non-Hermitian system and associated Hermitian systems is given by the time-dependent wave functions
$$
|\psi(t)\rangle =e^{-i t H}|\psi(0)\rangle \quad \mbox{and} \quad  |\phi(t)\rangle = S|\psi(t)\rangle = W\sqrt\eta|\psi(t)\rangle,
$$
respectively. The vector $|\psi(t)\rangle$ is normalized with respect to the metric induced by the inner product $\langle\cdot |\eta|\cdot\rangle$, while $|\phi(t)\rangle$ is normalized relative to the `original' inner product $\langle\cdot|\cdot\rangle$. The {\em reduced density matrices} for the non-Hermitian and the Hermitian systems are obtained by taking the partial trace over the bath degrees of freedom (see Section \ref{sec:bipartite}),
$$
\bar \rho_H(t) = {\rm tr}_\b \big( |\psi(t)\rangle\langle\psi(t)|\eta\big) \quad \mbox{and}\quad \bar\rho_{h_W}(t) = {\rm tr}_\b\big( |\phi(t)\rangle\langle\phi(t)|\big).
$$
Our main findings are summed up as follows:
\begin{itemize}
	
\item[1.] {\bf Explicit evolution.} We obtain explicit formulas for the states $|\psi(t)\rangle$ and $|\phi(t)\rangle$ as well as the reduced states $\bar\rho_H(t)$ and $\bar\rho_{h_W}(t)$. See Sections \ref{sec:reduced}, \ref{sec:reduced-hermitian}. 

\item[2.] {\bf Metric $\eta$.} The operator $\bar\rho_H(t)$ is a density matrix exactly when $\eta$ is of product form $\Lambda_\s\otimes\Lambda_\b$ -- otherwise $\bar\rho_H(t)$ has complex eigenvalues. See Section \ref{subs:qhs}. We thus take $\eta$ of product form in the further analysis.

\item[3.] {\bf Subsystem entropy.} The reduced states $\bar\rho_H(t)$ and $\bar\rho_{h_W}(t)$ are periodic\footnote{The whole $\s\b$ complex consists of $N+1$ oscillators, so the energy spectrum of all the Hamiltonians involved consists of discrete eigenvalues only, without continuous spectrum. This explains the periodicity.} in time, both having the same period regardless of the choice of $W$.  The von Neumann entropy $S=-{\rm tr}(\rho\ln\rho)$ of $\bar\rho_H(t)$ and $\bar\rho_{h_W}(t)$ is periodic in time as well, but for {\em generic initial conditions\footnote{Only for specially tuned initial states and $W$'s is the situation non-generic, see Section \ref{sec:reduced-hermitian}.} and generic $W$},
the period of the entropy of the Hermitian system is {\em double} that of the non-Hermitian system. See section \ref{sec:entropy}.

\item[4.] {\bf $\s\b$ entanglement.} The non-Hermitian and  the Hermitian $\s\b$ states $|\psi(t)\rangle$, $|\phi(t)\rangle$ are entangled for all times except at periodically reoccurring single instants.\footnote{For a class of exceptional initial conditions the states are entangled for {\em all} times, see Section \ref{sect:ent}.}
Given any entangled state $|\psi\rangle$, one can find $W$ such that the associated $|\phi\rangle$ is disentangled, and for any disentangled $|\psi\rangle$ there are $W$ such that $|\phi\rangle$ is entangled. However, in an averaged sense, the choice of $W$ does not influence the entanglement at all. Namely, the {\em concurrence} of the time-averaged density matrix, $\langle\rho\rangle = \frac1T\int_0^T |\phi(t)\rangle\langle\phi(t)|\,dt$, where $T$ is the period of $|\phi(t)\rangle\langle\phi(t)|$, is {\em independent} of $W$. Its value is determined entirely by the initial condition and the choice of the metric. We identify the initial states for which $\langle\rho\rangle$ is separable and for which it is maximally entangled. See Section~\ref{sect:ent}.
\end{itemize}

\section{Quasi-Hermitian systems}
\label{sec:qhs}

Let $\mathcal H$ be a finite-dimensional Hilbert space with  inner product $\langle \cdot |\cdot\rangle$. An operator $\eta$ is said to be positive, denoted as $\eta>0$, if $\langle\psi | \eta\psi\rangle>0$ for all nonzero $|\psi\rangle\in\h$. This is equivalent with saying that $\eta^\dag=\eta$ and all eigenvalues of $\eta$ are strictly positive. Here $A^\dag$ is the adjoint of the operator $A$, defined by $\langle \psi|A \phi\rangle = \langle A^\dag\psi|\phi\rangle$ for all $|\phi\rangle, |\psi\rangle \in\h$. An operator $H$ on $\h$ is called ($\eta$-){\em quasi-Hermitian} if there exists a positive operator $\eta>0$  such that 
\begin{equation}
	\label{m1}
	H^\dagger  = \eta H\eta^{-1}.
\end{equation}
Quasi-hermiticity is a special case of {\em pseudo-hermiticity}, where \eqref{m1} holds with an invertible (but not necessarily positive) Hermitian operator $\eta$. Pseudo- and quasi-hermitian Hamiltonians arise in $PT$-symmetric quantum theory, see for instance \cite{Zhang2020} and references therein. 

\subsection{Hermitian counterparts}

Let $H$ be a non-Hermitian operator on $\h$, a candidate for the Hamiltonian of a physical system.  In order to obtain a Hermitian quantum theory, one could either:
\begin{itemize}
\item[(1)] Modify the inner product of $\h$ to $\langle \cdot| \eta\, \cdot\rangle$ for some $\eta>0$ (called a metric operator), such that $H$ becomes Hermitian in the Hilbert space $\h_\eta$ with this new inner product; or
\item[(2)] Take a similarity transformation (invertible map) $S$ such that the transformed $h=SHS^{-1}$ is Hermitian in the original Hilbert space $\h$. 
\end{itemize}
If $H$ is quasi-Hermitian, then both options (1) and (2) are possible, but neither the metric nor the similarity transform in options (1) and (2) are unique. To explore this non-uniqueness, we first notice that any quasi-Hermitian $H$ is diagonalizable \cite{DH,FZ22,Mos2002I}. More precisely, 
\begin{equation}
	\label{m3}
	H = \sum_{n=1}^N E_n |\psi_n\rangle\langle\phi_n|,
\end{equation}
where the $E_n\in\rx$ are the eigenvalues and the $\{|\psi_n\rangle,|\phi_n\rangle\}_{n=1}^N$ form a complete bi-orthonormal family, meaning that $\langle\psi_k|\phi_l\rangle = \delta_{kl}$ and $\sum |\psi_n\rangle\langle\phi_n|=\bbbone$. Let us consider the case where all eigenvalues $E_n$ are distinct for simplicity (a discussion including degenerate eigenvalues can be done similarly, but this is not our focus here). Then the decomposition \eqref{m3} is unique, it is the spectral representation of the operator $H$, and the $P_n \equiv |\psi_n\rangle\langle\phi_n|$ are the uniquely defined (generally not orthogonal) spectral projections. The vectors $|\psi_n\rangle$ and $|\phi_n\rangle$, however, are determined only up to a joint scaling $|\phi_n\rangle\mapsto z_n|\phi_n\rangle$ and $|\psi_n\rangle \mapsto \frac{1}{\overline{z_n}} |\psi_n\rangle$, with $0\neq z_n\in\mathbb C$ arbitrary. 
\medskip

$\bullet$ First let us explore the option (1). A metric $\eta$ is called a {\em metric for $H$} if $H$ is $\eta$-quasi-Hermitian. Let $A$ be a linear operator on $\h$, with adjoint $A^\dag$ as defined above.  If $A$ is viewed as an operator on $\h_\eta$, then $\langle \phi| \eta A\psi\rangle = \langle \eta^{-1}A^\dag\eta\phi|\eta \psi\rangle$, so the adjoint of $A$ in $\h_\eta$ is $A^\ddag = \eta^{-1} A^\dag\eta$. It follows that  a given $\eta>0$ is a metric for $H$ if and only if $H^\ddag=H$, that is, if and only if $H$ is Hermitian acting on $\h_\eta$. It is well known (see e.g. \cite{Mos2010, FZ22}) that 
\begin{equation}
	\label{m4}
\eta \mbox{\ \, is a metric for $H$} \quad \Longleftrightarrow \quad
\eta = \sum_{n=1}^N x_n |\phi_n\rangle\langle\phi_n|\mbox{\quad  for some $x_1,\ldots,x_N>0$},
\end{equation}
where the $|\phi_n\rangle$ are the vectors appearing in \eqref{m4}. The multitude of metrics obtained by varying the $x_j$ in \eqref{m3} naturally appears due to the fact that $|\phi_n\rangle$ is only determined up to an arbitrary nonzero scaling factor $z_n$ (as explained after \eqref{m3}), which results in the scaling $x_n\mapsto x_n|z_n|^2$. Given this non-uniqueness of the metric, which one should be chosen to define the physical Hilbert space $\h_\eta$?

One answer is that the metric is fixed provided that instead of just $H$, one chooses a whole irreducible family of operators to be Hermitian observables. Namely, it is shown in \cite{SGH} (see also \cite{Mos2006} for the two-dimensional case) that if there is a family of operators $\{A_i\}_i$ on $\mathcal H$, and positive operators $\eta$, $\eta'$ such that $A_i^\dag=\eta A_i\eta^{-1}$ and $A_i^\dag=\eta' A_i(\eta')^{-1}$ for all $i$, then
$$
\mbox{$\eta'$ is a scalar multiple of $\eta$}\quad \Longleftrightarrow \quad \mbox{$\{A_i\}_i$ is an irreducible family of operators on $\mathcal H$.}
$$ 
This means that for an irreducible family of quasi-Hermitian operators, there is exactly one metric (up to a scalar multiple) that makes those operators Hermitian. The chosen family\footnote{Examples of irreducible families are the Pauli matrices for a spin, with the Euclidean inner product on ${\mathbb C}^2$, or the position $\hat x$ and momentum $\hat p=-i\hbar \nabla_x$ for a quantum particle (rather, the bounded Weyl operators generated by them) with the inner product $\langle\psi|\phi\rangle = \int_{{\mathbb R}^3}\bar\psi(x)\phi(x)\,d^3x$.} can then be viewed as the physical observables of the theory and the space of pure states is $\mathcal H$ with inner product $\langle\cdot|\cdot\rangle_\eta$.

On the other hand, if interested only in the single observable $H$ (the Hamiltonian), one should keep the $x_n$ in \eqref{m4} general. 
\medskip

$\bullet$ Next, let us investigate option (2) for $H$ of the form \eqref{m3}. 
Let $\eta$ be a metric for $H$, so it is of the form \eqref{m4}. We find all invertible $S$ such that the   transformed $h\equiv SHS^{-1}$ is Hermitian, 
\begin{equation}
	\label{2-4}
h = S H S^{-1} = \big( SH S^{-1}\big)^\dag = h^\dag.
\end{equation}
One readily sees that \eqref{2-4} is equivalent to $TH=HT$, where $T=\eta^{-1} S^\dag S$. That $T$ commutes with $H$, as in \eqref{m3}, is equivalent to $T$ being  diagonal in the same bi-orthonormal system as $H$, that is, $T=\sum_{n=1}^N t_n |\psi_n\rangle \langle \phi_n|$ for some $t_n\in\mathbb C$. Now
\begin{equation}
	\label{2-5}
S^\dag S=\eta T = \Big( \sum_{n=1}^N x_n |\phi_n\rangle\langle \phi_n|\Big) \Big(\sum_{k=1}^N t_k |\psi_k\rangle\langle \phi_k|\Big) = \sum_{n=1}^N x_nt_n |\phi_n\rangle\langle \phi_n|,
\end{equation}
and as $S^\dag S>0$ and $x_n>0$, we have $t_n>0$ as well. It follows from \eqref{m4} and \eqref{2-5} that $\eta T$ is also a metric for $H$. In fact, \eqref{m4} and \eqref{2-5} show that given a fixed metric $\eta$ for $H$ and varying $\eta T$ 
over all  operators $T>0$ that commute with $H$, we obtain all of the metrics for $H$. We conclude that given $\eta$, the $S$ we are looking for are  the solutions of $S^\dag S=\eta T$, where $T>0$ is an  operator that commutes with $H$ (equivalently, is diagonal in the same bi-orthonormal system as $H$). The general solution is 
\begin{equation}
\label{eq:SWeta}
S=W\sqrt{\eta T},
\end{equation}
where $W$ is any unitary and where for a positive operator $A$, $\sqrt{A}$ is the unique positive operator whose square equals $A$.

 Once $W$ and $T$ are chosen, the associated Hermitian $h$ in \eqref{2-4} becomes
\begin{equation}
	h_{W,T} = W \sqrt{\eta T}\, H\frac{1}{\sqrt{\eta T}} W^\dag.
\end{equation}
We stress with this notation that $h$ depends on the choice of $W$ and $T$. The $h$ obtained from two different choices of unitaries, say $V$ and $W$, are unitarily equivalent, with $h_{V,T} = U h_{W,T} U^\dag$ and  $U =VW^\dag$.
In this sense, the choice of $W$ is globally immaterial. However, if the Hilbert space has a local structure, say is of bipartite nature $\h=\h_\s\otimes\h_\b$, then the global unitary $U$ may well change the local properties of the two local subsystems, in which case the choice of $W$ will play a physically relevant role. We also point out that the spectrum of $h_{W,T}$ does not depend on either $W$ or $T$ (or $\eta$, for that matter).
\medskip

In this work we take the following approach: We start with a given quasi-Hermitian Hamiltonian $H$ and an arbitrary metric $\eta$ for $H$ and we view $\h_\eta$ as the physical Hilbert space. We analyze the class of all associated {\em Hermitian systems} $h_{W,T}$, where $W$ and $T$ vary over all unitaries and all positive operators commuting with $H$, respectively. As explained above, considering all metrics $\eta$ is the same as considering all metrics $\eta T$, so varying over $T$ is redundant if $\eta$ is kept arbitrary. We may then set $T=\bbbone$ and only consider
\begin{equation}
\label{m17}
S=W\sqrt\eta,\qquad \hW = W\sqrt{\eta}\,  H\frac{1}{\sqrt \eta} W^\dag
\end{equation}
for all $W$ and $\eta$.

\subsection{States, reduced states, von Neumann entropy}
\label{sec:bipartite}

Consider now a fixed metric $\eta$, so that the physical Hilbert space is $\h_\eta$ and $H$ is Hermitian on $\h_\eta$, $H^\ddag=H$. Then $e^{-i t H}$ is the unitary Schr\"odinger dynamics on $\h_\eta$. The average of an observable $A$ on $\h_\eta$ in the state $\psi\in\h_\eta$ is given by 
\begin{equation}
	\label{m5}
	\langle \psi|A\psi\rangle_\eta = \langle \psi| \eta A\psi\rangle = \tr\big(|\psi\rangle\langle\psi|\eta A\big) = \tr \big(\widetilde\rho A\big),
\end{equation}
where
\begin{equation}
	\label{6}
	\widetilde\rho = |\psi\rangle\langle\psi|\eta
\end{equation}
is a density matrix on $\h_\eta$ (a positive, trace-one operator). This $\widetilde\rho$ is called the `generalized density matrix' in \cite{ScSo}.
It is important to point out that the trace in \eqref{m5} is a purely algebraic quantity: it is the sum of the eigenvalues of the operator, and therefore does not depend on the choice of metric.

\bigskip

To arrive at a Hermitian Hamiltonian, it is necessary to make a choice for the unitary $W$ in \eqref{m17}. The  associated Hermitian Hamiltonian $\hW$ is then given by \eqref{m17}. Let
\begin{equation}
	|\psi(t)\rangle =e^{-i t H}|\psi(0)\rangle,\quad |\phi(t)\rangle =  e^{-i t \hW}|\phi(0)\rangle
\end{equation}
be the evolution of the initial states $|\psi(0)\rangle$, $|\phi(0)\rangle$ with respect to $H$ and $h_W$, respectively. The states are related by
\begin{equation}
	\label{m19}
	|\phi(t)\rangle = S|\psi(t)\rangle,\quad S=W\sqrt\eta,
\end{equation}
and the density matrices associated to these vector states for the non-Hermitian (see \eqref{6}) and the Hermitian systems are 
\begin{equation}
	\rho_{H}(t) = |\psi(t)\rangle\langle\psi(t)| \eta\quad\mbox{and}\quad \rho_{\hW}(t) = |\phi(t)\rangle\langle\phi(t)|,
\end{equation}
respectively.
(We adopt the notation $\rho_{\hW}$ and $\rho_H$ for the density matrices on the Hermitian and non-Hermitian sides of the problem from \cite{FF19}.) It is clear from \eqref{m19} that
\begin{equation}
	\label{m21}
	\rho_{{\hW}}(t) = S|\psi(t)\rangle\langle\psi(t)|S^\dag = S|\psi(t)\rangle\langle\psi(t)|(S^\dag S)S^{-1} = S\rho_H(t)S^{-1}.
\end{equation}
It follows that $\rho_{\hW}(t)$ and $\rho_H(t)$ have the same eigenvalues, and hence the same von Neumann entropy, $\e(\rho_{\hW}(t)) = \e(\rho_H(t))$, where
\begin{equation}
	\e(\rho) = -\tr\big(\rho\ln\rho\big) =  -\sum_{i}\lambda_i \ln\lambda_i
\end{equation}
and $\{\lambda_i\}$ are the eigenvalues of $\rho$.
\bigskip

Consider now a bipartite system with $\h=\h_\s\otimes\h_\b$  (`system' and `bath'). We consider the {\em reduced states} (denoted by an overbar) defined by tracing out the degrees of freedom of the subsystem $\h_\b$,
\begin{equation}
	\label{m22}
	\bar\rho_H(t) = \tr_{\h_\b} \big(\rho_H(t)\big),\qquad \bar\rho_{\hW}(t) = \tr_{\h_\b} \big(\rho_{\hW}(t)\big).
\end{equation}
In some recent works \cite{FF19, FrithPHD2020,Frith2020,Cen2020}, the dynamics of a bipartite system generated by a non-Hermitian Hamiltonian $H$ is studied, with particular focus on  the von Neumann entropy of the reduced density matrix $\bar\rho_H(t)$. The strategy proposed in those works is to examine the entropy of $\bar\rho_{\hW}(t)$ as a proxy for that of $\bar\rho_H(t)$. In this respect, however, one should observe the following facts:
\begin{itemize}
	\item[1.] The operator $\bar\rho_H(t)$ always satisfies $\tr_{\h_\s}(\bar\rho_H(t))=1$, but for some choices of $\eta$ the eigenvalues of $\bar\rho_H(t)$ can be complex, in which case it is not a valid density matrix.
	
	\item[2.] Even if the metric $\eta$ is chosen such that $\bar\rho_H(t)$ is a density matrix, for generic choices of $W$ the von Neumann entropies $\e(\bar\rho_H(t))$ and $\e(\bar\rho_{\hW}(t))$ are not the same. The latter in fact depends on the choice of $W$.
\end{itemize}

To understand the normalization of the trace mentioned in fact 1.~above, we observe (using $\bbbone_\s$ as the system observable) that
$$
\tr_{\h_\s}(\bar\rho_H(t)) = \tr_{\h_\s}(\bar\rho_H(t)\bbbone_\s)=\tr_{\h_\s\otimes\h_\b}\big(\rho_H(t) (\bbbone_\s\otimes\bbbone_\b) \big)=\tr_{\h_\s\otimes\h_\b}(\rho_H(t))=1.
$$
If $S=S_\s\otimes S_\b$, then $\bar\rho_h=S_\s \bar \rho_HS_\s^{-1}$ and so the spectra and thus the von Neumann entropies of $\bar\rho_h$ and $\bar\rho_H$ coincide. However, if $S$ is entangling (not of product form $S_\s\otimes S_\b$), then the eigenvalues of the two reduced density matrices are not the same in general, and neither are their entropies. 

These difficulties are resolved in the next section, where we study the concrete model used in \cite{FF19}. In particular, we determine for which choices of $\eta$ the reduced operator $\bar\rho_H(t)$ is indeed a density matrix, and then we find the von Neumann entropy of $\bar\rho_{h_W}(t)$ {\em for all possible choices of the unitary $W$}.

\section{Model}
\label{sec:model}

An oscillator with creation and annihilation operators $a^\dagger$, $a$ is coupled to a `bath' of $N$ independent oscillators with creation and annihilation operators $q_i^\dagger$, $q_i$, $i=1,\ldots,N$.
The total Hilbert space of the $N+1$ oscillators is
\begin{equation}
	\label{Htot}
\mathcal H = \mathcal H_\s\otimes\mathcal H_\b,
\end{equation}
where $\mathcal H_\s$ is the space of a single oscillator and $\mathcal H_\b$ is that of the other $N$. As in the previous section, we denote the inner product by $\langle\cdot|\cdot\rangle$ and let
$\dagger$ denote the adjoint in this inner product.
The commutation relations are $[a,a^\dagger]=1=[q_i,q^\dagger_i]$, and all operators belonging to different oscillators commute. This open quantum system, non-Hermitian model was used in \cite{FF19}.

\subsection{The quasi-Hermitian system}
\label{subs:qhs}

The coupled total system--bath Hamiltonian is 
\begin{equation}
\label{1}
H =\nu \N +(g+\kappa)\sqrt N \, a^\dagger Q + (g-\kappa)\sqrt N\ a Q^\dagger,
\end{equation}
where $\nu>0$ and $g,\kappa\in \mathbb R$ are parameters and 
\begin{equation}
\N = a^\dagger a + \sum_{n=1}^N q_n^\dagger q_n,\qquad Q=\frac{1}{\sqrt N}\sum_{n=1}^N q_n.
\end{equation}
Due to the different prefactors of $\kappa$ in the interaction term of \eqref{1}, $H$ is $\dagger$-Hermitian if and only if $\kappa= 0$.

The `uncoupled' ($g=\kappa=0$) Hamiltonian is simply $\nu \N$, a multiple of the total number operator $\N$. As $H$ commutes with $\N$, each eigenspace of $\N$, with a fixed number of excitations (in the system plus the bath) is left invariant. Denote by $|0_\s 0_\b\rangle$ the `vacuum' zero excitation state, where all oscillators are in the ground state. The single excitation space is defined as
\begin{equation}
{\mathcal E}_1 = {\rm span} \big\{ |1_\s 0_\b\rangle, |0_\s 1_1\rangle, |0_\s 1_2\rangle,\ldots, |0_\s 1_N\rangle \big\},
\end{equation}
where $|1_\s 0_\b\rangle = a^\dagger |0_\s 0_\b\rangle$  and $|0_\s 1_i\rangle = q^\dagger_i |0_\s 0_\b\rangle$ for $i=1,\ldots,N$. When $H$ is applied to a vector in ${\mathcal E}_1$ the result is again a vector in ${\mathcal E}_1$. Moreover, due to the collective, symmetric nature of the system-bath interaction in \eqref{1}, $H$ leaves the even smaller space 
\begin{equation}
\label{m28}
{\mathcal H}_1 = {\rm span} \big\{ |e_\s\rangle,\ |e_\b\rangle\big\}
\end{equation}
invariant, where 
\begin{equation}
|e_\s\rangle=|1_\s 0_\b\rangle,\qquad |e_\b\rangle= \frac{1}{\sqrt N} \sum_{n=1}^N |0_\s 1_n\rangle.
\label{m28.1}
\end{equation}
Those two vectors describe states in which a single excitation is either in $\s$ (the state $|e_\s\rangle$) or in $\b$, collectively spread over the $N$ bath oscillators (the state $|e_\b\rangle$). 
Therefore, we may view $H$ as an operator on ${\mathcal H}_1$. When we do this we denote it by $H_1$, which has the form
\begin{equation}
\label{4}
H_1 = \nu \bbbone + (g-\kappa) \sqrt N \ |e_\b\rangle\langle e_\s| + (g+\kappa) \sqrt N \ | e_\s\rangle \langle e_\b|.
\end{equation}
The eigenvalues of $H_1$ are 
\begin{equation}
\omega_\pm = \nu \pm \omega,\qquad \omega=\sqrt N \sqrt{g^2-\kappa^2},
\end{equation}
which are real for $\kappa^2\le g^2$ and (purely imaginary) complex conjugates for $\kappa^2>g^2$. See \cite{FF19} for a discussion of the $PT$ symmetry of $H$. The operator $H_1$ is diagonalizable except at the transition points defined by $\kappa^2 = g^2 \neq 0$, where $H_1$ reduces to a Jordan block. Note that increasing the number $N$ of oscillators in the bath simply amounts to speeding up the dynamics (the frequency $\omega$) by a factor $\sqrt N$.  

We consider the `$PT$-symmetry unbroken regime' $\kappa^2<g^2$, so that $\omega_\pm \in \rx$. For definiteness we take $g>0$ (the case $g<0$ can be dealt with in the same fashion), so
\begin{equation}
\label{regime}
0\le |\kappa|< g,
\end{equation}
which is equivalent to $g+\kappa>0$ and $g-\kappa>0$. Then we have $\omega>0$  and 
\begin{equation}
\label{a}
a_1=\sqrt{g+\kappa}>0,\quad a_2=\sqrt{g-\kappa}>0,
\end{equation} 
where the equalities in \eqref{a} define the quantities $a_1$, $a_2$. 
The two linearly independent (not normalized) eigenvectors of $H_1$  and its adjoint $H_1^\dagger$ are 
\begin{equation*}
|v_\pm\rangle \propto a_1 |e_\s\rangle \pm a_2 |e_\b\rangle \quad\mbox{and}\quad |v^*_\pm\rangle \propto a_2 |e_\s\rangle \pm a_1 |e_\b\rangle,
\end{equation*}
respectively. They satisfy $H_1|v_\pm\rangle = \omega_\pm|v_\pm\rangle$ and $H^\dagger_1|v^*_\pm\rangle = \omega_\pm |v^*_\pm\rangle$. Note that $|v_\pm^*\rangle$ denote the eigenvectors of $H^\dag$, not to be confused with the complex conjugates of the eigenvectors $|v_\pm\rangle$  of $H$. We normalize the vectors as
\begin{equation}
	\label{36}
|v_\pm\rangle =\tfrac{1}{\sqrt 2} \Big( \sqrt{\tfrac{a_1}{a_2}} |e_\s\rangle \pm \sqrt{\tfrac{a_2}{a_1}} |e_\b\rangle\Big)  \quad\mbox{and}\quad |v^*_\pm\rangle =\tfrac{1}{\sqrt 2}\Big(  \sqrt{\tfrac{a_2}{a_1}} |e_\s\rangle \pm \sqrt{\tfrac{a_1}{a_2}} |e_\b\rangle\Big).
\end{equation}
Then $\{ |v_\pm\rangle ,|v^*_\pm\rangle\}$ is a bi-orthonormal basis, satisfying $\langle v^*_\pm| v_{\mp}\rangle=0$ and $\langle v^*_\pm|v_\pm\rangle=1$, and the operator $H_1$ can be written as
\begin{equation}
\label{19.1}
H_1 = \omega_+ |v_+\rangle\langle v^*_+| +\omega_- |v_-\rangle\langle v^*_-|.
\end{equation}
Using this, one easily finds
\begin{eqnarray}
e^{-i t H_1} &=& e^{-i t \omega_+} |v_+\rangle\langle v^*_+| + e^{-i t \omega_-} |v_-\rangle\langle v^*_-| \nonumber \\
&=& e^{- i t\nu}\cos(\omega t)\bbbone -ie^{-i t \nu} \sin(\omega t) \Big( \frac{a_1}{a_2}|e_\s\rangle\langle e_\b|+ \frac{a_2}{a_1}|e_\b\rangle\langle e_\s|\Big).
\end{eqnarray}

We consider initial states which are vectors in ${\mathcal H}_1$, as defined in \eqref{m28}, so the dynamics generated by $H$ is entirely given by the operator $H_1$ from \eqref{4}. We still consider the regime \eqref{regime}, so that the spectrum of $H_1$  consists of two distinct real eigenvalues. Comparing \eqref{m3}, \eqref{m4} and \eqref{19.1}, we see that $H_1$ is quasi-Hermitian and the set of all associated metrics is
\begin{equation}
	\label{11}
	{\mathcal M}_+=\big\{\eta= x_1 |v_+^*\rangle\langle v_+^*| +x_2 |v_-^*\rangle\langle v_-^* | \ :\ x_1,x_2>0 \big\}.
\end{equation}
Written as a matrix in the basis $\{|e_\s\rangle, |e_\b\rangle \}$, we obtain from \eqref{36}
\begin{equation}
	\label{metricmatrix-1}
	\eta =\frac12
	\begin{pmatrix}
		(x_1+x_2)\, a_2/a_1 & x_1-x_2\\
		x_1-x_2 & (x_1+x_2) \, a_1/a_2
	\end{pmatrix}.
\end{equation}
This is diagonal exactly when $x_1=x_2$. As we will see in Section \ref{sect3.4}, this is equivalent to $\eta$ being the {\em restriction to $\h_1$ of a  product metric $\Lambda_\s\otimes\Lambda_\b$ on $\h$}. And as we discuss below after \eqref{19}, this is also equivalent to the reduced system state $\bar\rho_H(t)$ in \eqref{53} being a positive operator.

\subsection{Reduced non-Hermitian system dynamics}
\label{sec:reduced}

Fix an $\eta\in\mathcal M_+$ and take an initial state of the form 
\begin{equation}
\label{ic}
|\psi(0)\rangle = A |e_\s\rangle + B |e_\b\rangle
\end{equation}
for some $A,B\in\mathbb C$ normalized to have $\|\psi(0)\|^2_\eta=1$, that is,
\begin{equation}
\label{m38}
1= \Big( \frac{x_1+x_2}{2}\Big) \Big( \frac{a_2}{a_1} |A|^2 +\frac{a_1}{a_2}|B|^2\Big) +(x_1-x_2){\rm Re} ( A  B^*).
\end{equation}
The dynamics is given by
\begin{equation}
\label{12}
|\psi(t)\rangle = e^{- i t H} |\psi(0)\rangle =e^{-it\nu}A(t)  |e_\s\rangle + e^{-i t\nu} B(t) |e_\b\rangle,
\end{equation}
where
\begin{equation}
\begin{aligned}
A(t) &= A\cos(\omega t) -iB\frac{a_1}{a_2}\sin(\omega t)\label{m39},\\
B(t) &= B\cos(\omega t) -i A\frac{a_2}{a_1}\sin(\omega t).
\end{aligned}
\end{equation}
The normalization 
\begin{equation}
	\label{12.1}
\|\psi(t)\|^2_\eta=	x_1 \big|\langle v_+^*|\psi(t)\rangle\big|^2 + x_2 \big|\langle v_-^*|\psi(t)\rangle\big|^2 =1
\end{equation}
holds for all $t$, as $e^{-i t H}$ acts unitarily on ${\mathcal H}_1$ equipped with the inner product $\langle \cdot| \cdot\rangle_\eta$. The relation \eqref{12.1} is the same as \eqref{m38} with $A$ and $B$ replaced by $A(t)$ and $B(t)$.
\bigskip

We now introduce the reduction of the system to the single oscillator ($a^\dag$, $a$). 
The average of a system observable $O_\s$ (observable of the single oscillator) in the state $|\psi(t)\rangle$ given by \eqref{12} evolves according to 
\begin{equation}
\langle \psi(t)|\eta O_\s|\psi(t)\rangle = \tr_\s \big(\bar\rho_H(t) O_\s\big),
\end{equation}
where the reduced system state is
\begin{equation}
	\label{53}
\bar\rho_H(t) ={\rm tr}_\b \, \rho_H(t) = \tr_\b \big( |\psi(t)\rangle\langle\psi(t)| \eta\big).
\end{equation}
For the partial trace we have the identities ${\rm tr}_\b |e_\s\rangle\langle e_\s| = |1_\s\rangle\langle 1_\s|$, ${\rm tr}_\b |e_\b\rangle\langle e_\b| = |0_\s\rangle\langle 0_\s|$ and 
${\rm tr}_\b |e_\s\rangle\langle e_\b| = 0 = {\rm tr}_\b |e_\b\rangle\langle e_\s|$. 
Using \eqref{12} and $\eta$ of the form \eqref{11}, we obtain after a calculation
\begin{eqnarray}
	\label{19}
\bar\rho_H(t) &=& \Big( \frac{x_1+x_2}{2} \frac{a_1}{a_2}|B(t)|^2 +\frac{x_1-x_2}{2} A(t)B(t)^*\Big)  \, |0_\s\rangle\langle 0_\s|\nonumber\\
&&+\Big( \frac{x_1+x_2}{2} \frac{a_2}{a_1}|A(t)|^2 +\frac{x_1-x_2}{2} A(t)^* B(t)\Big)  \, |1_\s\rangle\langle 1_\s|.
\end{eqnarray}
This matrix is diagonal in the basis $\{|0_\s\rangle, |1_\s\rangle\}$ and the two diagonal entries are its eigenvalues. One checks directly that $\tr_\s( \bar\rho_H(t))=1$ (the sum of the diagonal elements equals $\|\psi(t)\|_\eta^2=1$). However, the eigenvalues of $\bar\rho_H(t)$ are complex, in general, unless the metric is chosen to satisfy $x_1=x_2$. Indeed, the imaginary part of the first eigenvalue is $\frac{x_1-x_2}{2} {\rm Im}  ( A(t)B(t)^*)$. If $A,B$, the coefficients in the initial state \eqref{ic}, are real then this quantity becomes\footnote{Use the equations \eqref{m39} and write $B$ as a function of $A$ according to the normalization condition \eqref{m38}.}  
$ - (x_1-x_2)(\frac{1}{x_1+x_2}-\frac{a_2}{a_1}A^2 - \frac{x_1-x_2}{x_1+x_2} AB) \cos(\omega t)\sin(\omega t)
$.
Unless $x_1=x_2$ or the initial condition satisfies $(x_1+x_2)\frac{a_2}{a_1}A^2 +(x_1-x_2)AB=1$, the eigenvalues of $\bar\rho_H(t)$ will not be real except at the discrete set of times $t$ when $\sin(\omega t)\cos(\omega t)=0$.

We require $\bar\rho_H(t)$ to be a density matrix (and in particular to have non-negative eigenvalues) for all times. To do so with a metric that does not depend on the initial conditions we therefore must choose $x_1 = x_2$. We thus take 
$$
x_1=x_2=x >0
$$ 
for the remainder of the paper. In the basis $\{|e_\s\rangle, |e_\b\rangle \}$ the metric $\eta$ is diagonal,
\begin{equation}
	\label{metricmatrix}
\eta = x
\begin{pmatrix}
a_2/a_1 & 0\\
0 & a_1/a_2
\end{pmatrix},
\end{equation}
see \eqref{metricmatrix-1}. As explained after \eqref{metricmatrix-1} above, this is equivalent to $\eta$ being of product form. With this choice, $\bar\rho_H(t)$ given by \eqref{19}  is $\dag$-Hermitian. According to \eqref{19} and \eqref{m39} we have 
\begin{equation}
\label{m45}
\bar\rho_H(t) = p(t) \,|0_\s\rangle\langle 0_\s| +(1-p(t)) \,|1_\s\rangle\langle 1_\s|,
\end{equation}
where
\begin{eqnarray}
p(t) &=& x\frac{a_1}{a_2}|B(t)|^2 \nonumber\\
&=& x \Big(\frac{a_1}{a_2} |B|^2\cos^2(\omega t) +\frac{a_2}{a_1}|A|^2\sin^2(\omega t) -2\sin(\omega t)\cos(\omega t) {\rm Im}(A^*B)\Big) \nonumber\\
&=& \frac12 + \Big(\frac12-x\frac{a_2}{a_1}|A|^2\Big)\cos(2\omega t) - x\sin(2\omega t){\rm Im}(A^*B).
\label{2-49}
\end{eqnarray}
In the last step, we used the normalization condition \eqref{m38}, resulting in $\frac{a_1}{a_2}|B|^2=\frac{1}{x} -\frac{a_2}{a_1}|A|^2$, and the trigonometric identities $\sin(\omega t)\cos(\omega t)=\frac12\sin(2\omega t)$, $\cos^2(\omega t) = \frac12 (1+\cos(2\omega t))$ and $\sin^2(\omega t)=\frac12(1-\cos(2\omega t))$. In view of \eqref{2-49} it is natural to introduce the parameter
\begin{equation}
	\label{alpha}
	\alpha \equiv x\frac{a_2}{a_1}|A|^2\in [0,1].
\end{equation}
Equations \eqref{m45} and \eqref{2-49} show the following.
\medskip

\noindent {\bf Properties of $p(t)$:}
\begin{itemize}
\item[1.] $p(t)$  and  $\bar\rho_H(t)$  depend on time unless $\alpha=\tfrac12$ and $A^*B\in\mathbb R$, in which case  $p(t)=\tfrac12$ and $\bar\rho_H(t)=\tfrac12\bbbone$.
\item[2.] Otherwise $p(t)$ and $\bar\rho_H(t)$ are periodic in time, with period $\pi/\omega$, and the  mean value of $p(t)$ is 
\begin{equation}
	\label{p0}
p_0 =\frac\omega\pi \int_0^{\pi/\omega}p(t)dt = \frac12.
\end{equation}
\end{itemize}

\subsection{Reduced Hermitian system dynamics}
\label{sec:reduced-hermitian}

Next, we turn our attention to the density matrix of the Hermitian system, which according to \eqref{m21} is
\begin{equation}
\label{24}
\rho_{\hW}(t) = S\rho_H(t) S^{-1} = W \sqrt{\eta} \rho_H(t)\tfrac{1}{\sqrt{\eta}} W^\dag = W\sqrt{\eta}\, |\psi(t)\rangle\langle\psi(t)|\sqrt{\eta}\, W^\dag.
\end{equation}
We keep $W$ in the notation $\hW$ to highlight that the choice of $h$ depends on $W$, see \eqref{m17}. Again choosing a metric $\eta$ of the form \eqref{11} with $x_1=x_2=x>0$, we use \eqref{12} to obtain
\begin{equation}
\sqrt{\eta}|\psi(t)\rangle =e^{-i t\nu}  \gamma(t) |e_\s\rangle + e^{-i t\nu}\delta(t)|e_\b\rangle,
\end{equation}
where 
\begin{equation}
 \gamma(t) = \sqrt{x\frac{a_2}{a_1}} A(t),\qquad 
\delta(t) =   \sqrt{x\frac{a_1}{a_2}} B(t).
\label{m49}
\end{equation}
We then obtain
\begin{eqnarray}
	\label{25}
\sqrt{\eta}\, |\psi(t)\rangle\langle\psi(t)|\sqrt{\eta} =
\begin{pmatrix}
|\gamma(t)|^2 &\gamma(t) \delta(t)^*\\
\gamma(t)^* \delta(t)& |\delta(t)|^2
\end{pmatrix},
\end{eqnarray}
written in matrix form in the ordered basis $\{|e_\s\rangle,|e_\b\rangle \}$ of $\h_1$.
Next, we take a general (time-independent) unitary on $\h_1$, expressed in the same basis as
\begin{equation}
	\label{26}
W = \begin{pmatrix}
a & b\\
c & d
\end{pmatrix},\qquad a c^*+bd^*=0,\quad |a|^2+|b|^2=1= |c|^2+|d|^2.
\end{equation}
Using \eqref{24}, \eqref{25}, \eqref{26} and writing momentarily $\delta,\gamma$ for $\delta(t), \gamma(t)$, we get
\begin{equation}
	\label{rhwmatrix}
\rho_{\hW}(t) =
\begin{pmatrix}
 |a\gamma+b\delta|^2 \quad & ac^*|\gamma|^2 + bc^*\gamma^*\delta + a d^* \gamma \delta^* +b d^*|\delta|^2\\
a^* c|\gamma|^2+b^* c \gamma \delta^* +a^* d \gamma^*\delta  +b^* d |\delta|^2 & |c\gamma +d\delta|^2
\end{pmatrix}.
\end{equation}
We observe that $\rho_{h_W}(t)$ is periodic in time, with period $\pi/\omega$. This follows from \eqref{m49} together with \eqref{m39}, since $\cos^2(\omega t)$, $\sin^2(\omega t)$ and $\sin(\omega t)\cos(\omega t)$ all have period $\pi/\omega$.


Next we calculate the reduced density matrix $\bar\rho_{\hW}(t)$ of $\s$ by taking the partial trace of $\rho_{\hW}(t)$ over $\b$, 
\begin{equation}
\label{m56}
\bar\rho_{\hW}(t)  =   q(t)  \, |0_\s\rangle\langle 0_\s| +\big(1-q(t)\big) \, |1_\s\rangle\langle 1_\s|,
\end{equation}
where $q(t) = \big| c\gamma(t)+d\delta(t) \big|^2$ and $\gamma(t), \delta(t)$ are given in \eqref{m49}. We get the expression
\begin{equation}
\label{m57}
q(t) = x\Big| c\sqrt{\frac{a_2}{a_1}} A(t) + d \sqrt{\frac{a_1}{a_2}}B(t)\Big|^2,
\end{equation}
where $A(t)$ and $B(t)$ are given in \eqref{m39}. Expanding the square and using $|d|^2=1-|c|^2$ as well as the normalization \eqref{m38} (which is valid for $A, B$ replaced by $A(t), B(t)$ at any time), we obtain
\begin{eqnarray}
	q(t) &=&|c|^2 +(1-2|c|^2) x
	\frac{a_1}{a_2}  |B(t)|^2+2x{\rm Re} \big(c d^* A(t) B(t)^*\big)\nonumber\\
	&=&|c|^2 +(1-2|c|^2) p(t) +2x{\rm Re} \big(c d^* A(t) B(t)^*\big),
\label{2-60}
\end{eqnarray}
where $p(t)$ is the population of $\bar\rho_H(t)$ evaluated above in \eqref{2-49}. Expanding the real part term in \eqref{2-60} using \eqref{m39}, we arrive at
\begin{eqnarray}
	q(t) &=&|c|^2 +(1-2|c|^2) p(t)\nonumber\\
	&&- \sin(2\omega t) \big(1-2x\frac{a_2}{a_1}|A|^2\big){\rm Im}(cd^*) -2x \cos(2\omega t) {\rm Im }(AB^*){\rm Im}(cd^*)\nonumber\\
	&&+2x{\rm Re}(AB^*){\rm Re}(c d^*).
	\label{2-61}
\end{eqnarray}
We take into account \eqref{2-49} to rewrite
\begin{eqnarray}
q(t) &=& \tfrac12 +2x {\rm Re}(AB^*){\rm Re}(cd^*)\nonumber\\
&&+ \cos(2\omega t)\Big[ (\tfrac12 -\alpha) (1-2|c|^2) -2x {\rm Im}(AB^*){\rm Im}(cd^*)\Big]\nonumber\\
&&- \sin(2\omega t)\Big[(1-2\alpha){\rm Im}(cd^*) - x(1-2|c|^2) {\rm Im}(AB^*)\Big],
\label{2-62}
\end{eqnarray}
where we recall that $\alpha$ is given in \eqref{alpha}. Using the explicit form \eqref{2-62} of $q(t)$, we obtain the following information.
\medskip

\noindent {\bf Properties of $q(t)$:}
\begin{itemize}
\item[1.] $q(t)$  and  $\bar\rho_{h_W}(t)$  depend on time unless both factors of the cosine and sine terms in \eqref{2-62} vanish. Thus $q(t)$ is time-independent and hence equal to 
\begin{equation}
	\label{q0}
q_0 =\frac\omega\pi \int_0^{\pi/\omega}q(t)dt = \frac12+2x{\rm Re} (AB^*){\rm Re}(cd^*)
\end{equation}
if and only if:
\begin{itemize}
    \item [$\bullet$] $\alpha=1/2$, $|c|=1/\sqrt2$ and ${\rm Im}(AB^*){\rm Im}(cd^*)=0$, or
    \item[$\bullet$] $\alpha=1/2$, $|c|\neq 1/\sqrt 2$ and ${\rm Im}(AB^*)=0$, or
    \item[$\bullet$] $\alpha\neq 1/2$, $|c|=1/\sqrt2$ and ${\rm Im}(cd^*)=0$, or
    \item[$\bullet$] $\alpha\neq 1/2$, $|c|\neq 1/\sqrt2$ and ${\rm Im}(A^*B) = \frac{1}{2x}|1-2\alpha|$ and ${\rm Im}(cd^*)=\frac12 (1-2|c|^2){\rm sgn}(1-2\alpha)$, where ${\rm sgn}(x)=|x|/x$.
\end{itemize}

\color{black}
\item[2.] Otherwise $q(t)$ and $\bar\rho_{h_W}(t)$ are periodic in time, with period $\pi/\omega$, and the  mean value of $q(t)$ is $q_0$ from \eqref{q0}.

\end{itemize}
 
\medskip

\noindent {\bf Generic initial states and unitaries.\ } As $x>0$, the average $q_0$ equals $\frac12$ exactly when ${\rm Re} (AB^*){\rm Re}(cd^*)=0$. This is a condition on the initial state (via $A,B$) and the unitary $W$ (via $c,d$). We call the initial state and the unitary {\em generic}, respectively, when 
\begin{equation}
\label{generic}
\mbox{$AB^*\not\in\mathbb R$ \quad and\quad   $cd^*\not\in\mathbb R$}.
\end{equation}
In other words, for generic initial states and unitaries, the average $q_0$ of the population of $\bar\rho_{h_W}(t)$  differs from the average $p_0=\frac12$ of the population of $\bar\rho_H(t)$.   As we show in the next section, this deviation from the value $\frac12$ causes the entropy of $\bar\rho_{h_W}(t)$ to oscillate with exactly {\em half} the frequency of the entropy of $\bar \rho_H(t)$.

\section{Entropy}
\label{sec:entropy}

Recall that the states $\bar\rho_{H}(t)$ and $\bar\rho_{\hW}(t)$ are given in \eqref{m45} and \eqref{m56}, with associated populations $p(t)$, $q(t)$ evaluated in \eqref{2-49} and \eqref{2-62}. Their von Neumann entropy is given by
\begin{eqnarray}
\mathcal E(\bar\rho_H(t))&=& -p(t)\ln p(t) - (1-p(t))\ln(1-p(t)),\nonumber \\
\mathcal E(\bar\rho_{\hW}(t))&=& -q(t)\ln q(t) - (1-q(t))\ln(1-q(t)).
\label{2020}
\end{eqnarray}
We show in Appendix \ref{AppB} that 
$$
\mathcal E(\bar\rho_H(t)) = \mathcal E(\bar\rho_{\hW}(t)) \mbox{\ for all $t\ge 0$ and initial conditions ($A,B$)\,} \quad \Longleftrightarrow\quad  cd=0.
$$ 
If ${\rm Re}(cd^*)\neq 0$, then according to \eqref{p0} and \eqref{q0} the averages $p_0$ and $q_0$ around which the populations $p(t)$ and $q(t)$ oscillate are {\em different} for all generic initial conditions, i.e., all coefficients $A,B$ satisfying  ${\rm Re}(AB^*)\neq 0$. This translates into a modification of the period of the entropy of $\bar\rho_{h_W}(t)$ as a function of time $t$ relative to that of $\bar\rho_H(t)$, as we explain now.

\subsection{Period doubling of  von Neumann entropy in the non-Hermitian versus the Hermitian system}

\begin{figure}[H]
	\centering
	\begin{subfigure}{.5\textwidth}
		\centering
		\includegraphics[width=1.2\linewidth]{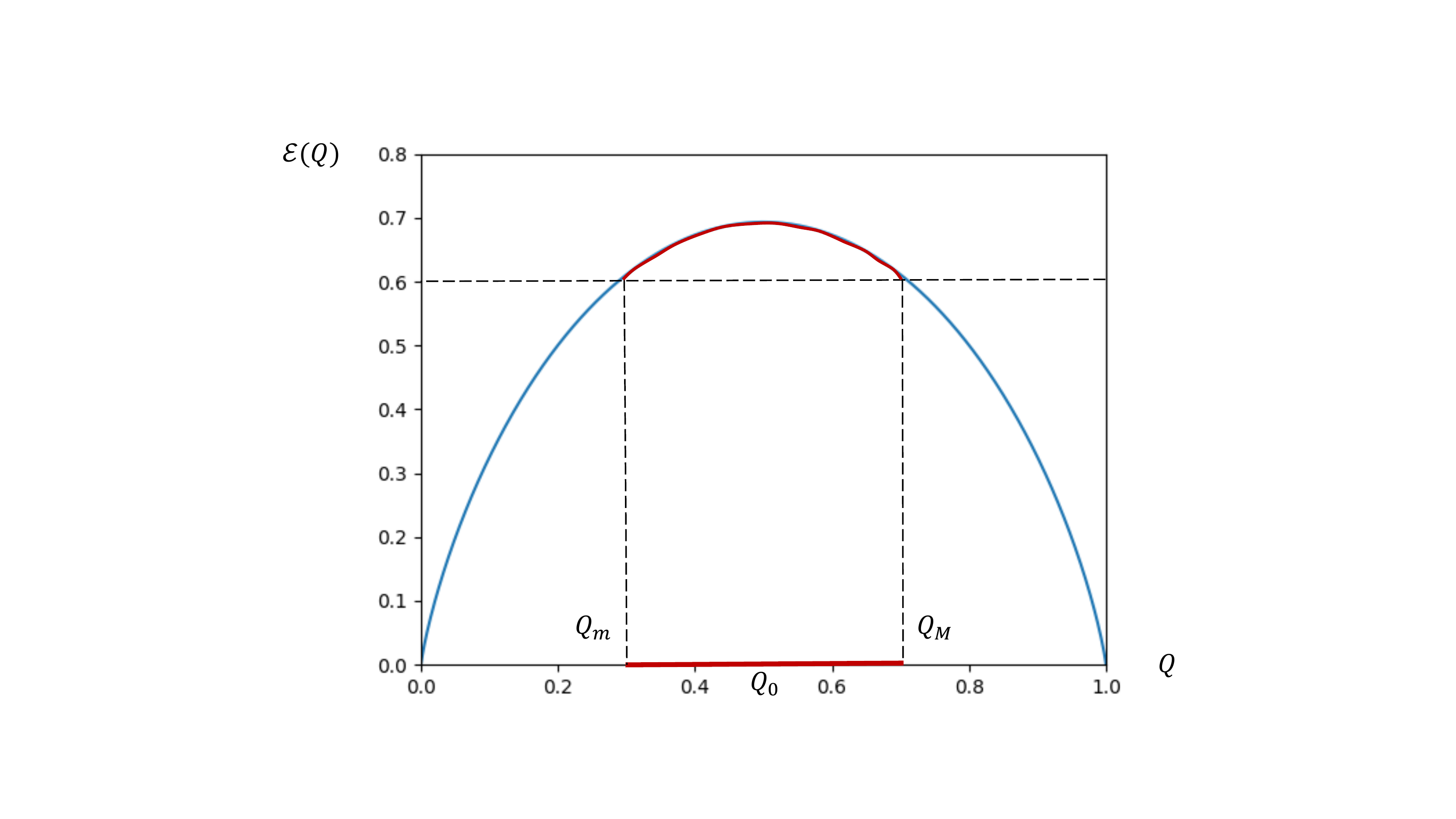}
  \vspace*{-1cm}
	\end{subfigure}%
	\begin{subfigure}{.5\textwidth}
		\centering
  \includegraphics[width=.75\linewidth]{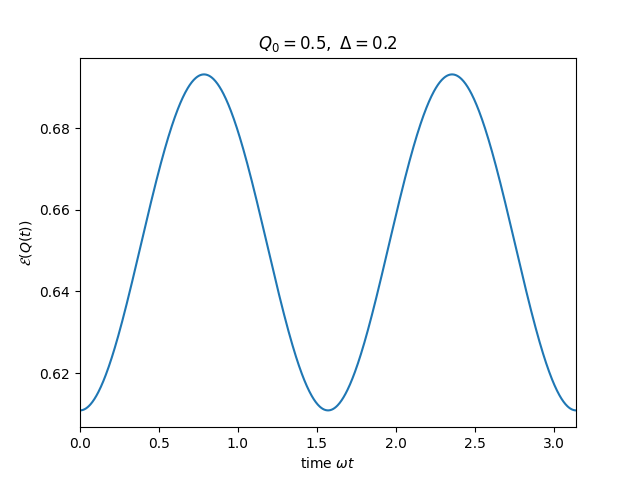}
	\end{subfigure}
	\caption{Parameters $Q_0=0.5$, $\Delta\equiv  \max_t\Delta(t)=0.2$. According to \eqref{2-Q}, $Q(t)$ starts at $Q_m=0.3$ (at a time we take to be $t=0$) and moves to $Q_M=0.7$ at time $\omega t=\pi/2$, and then back to $Q_m$ at time $\omega t=\pi$ (left panel), so the value of the entropy $\mathcal E(Q(t))$ evolves through two periods (right panel). In each period, the entropy has two local minima (counting minima at the endpoints of the considered intervals once). 
	}
	\label{fig:test}
\end{figure}

Consider the function
$$
\mathcal E(Q) =- Q\ln Q-(1-Q)\ln(1-Q),\quad Q\in[0,1].
$$
Suppose now that $Q=Q(t)$ depends periodically on time and has average $Q_0$, 
\begin{equation}
	\label{2-Q}
Q(t) =Q_0 + \Delta(t) \in [0,1],
\end{equation}
with $\Delta(t)$ having period $\pi/\omega$ and zero average. This setup incorporates both cases $p(t)$ and $q(t)$ in one. As Figure~\ref{fig:test} illustrates, if $Q_0=\frac12$, which is the value where $\mathcal E(Q)$ takes its maximum, then as $Q(t)$ moves over one period, the entropy $\mathcal E(Q(t))$ moves over {\em two} periods.

\begin{figure}[H]
	\centering
	\begin{subfigure}{.5\textwidth}
		\centering
		\includegraphics[width=1.2\linewidth]{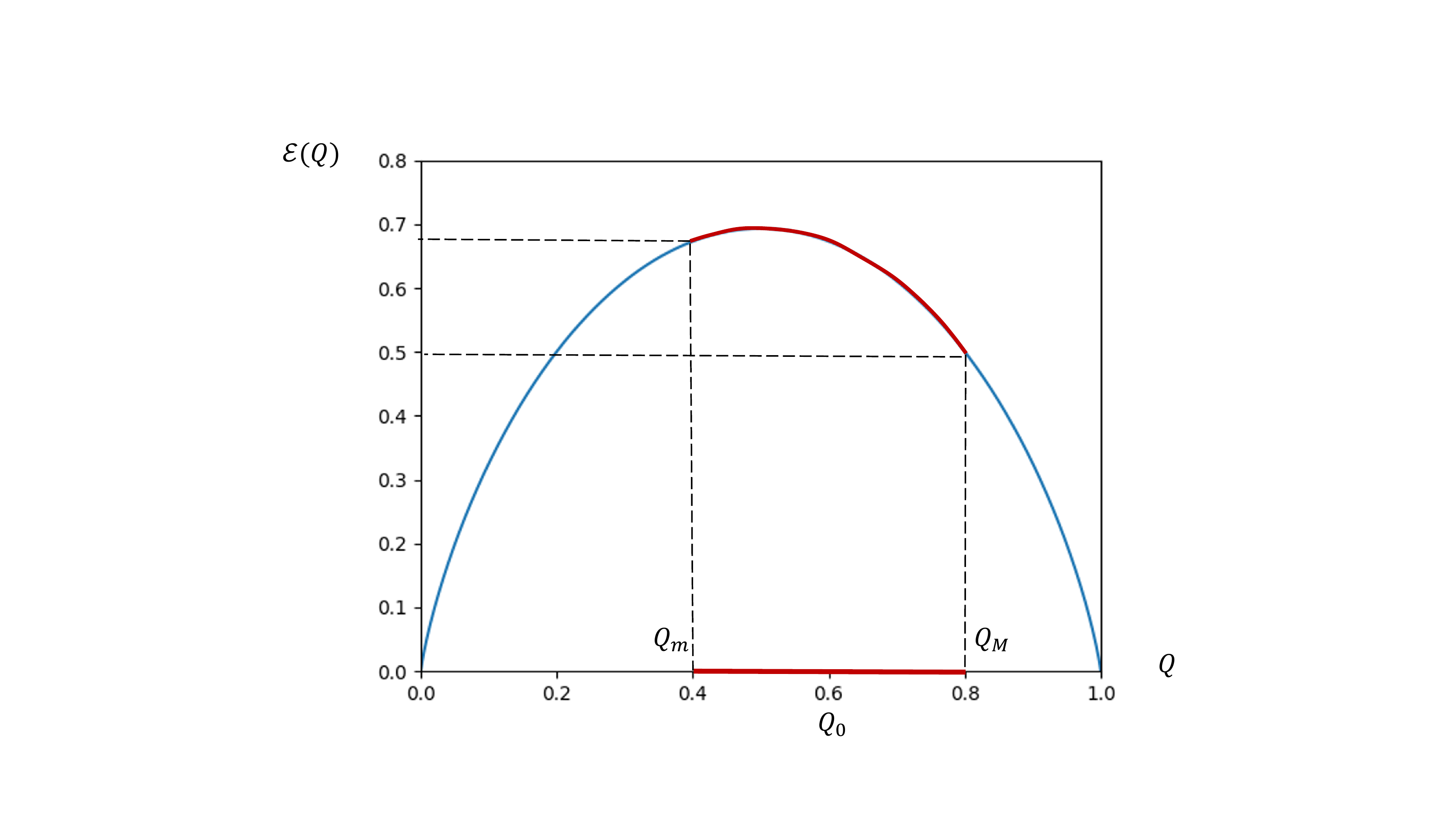}
  \vspace*{-1cm}
	\end{subfigure}%
	\begin{subfigure}{.5\textwidth}
		\centering
		\includegraphics[width=.75\linewidth]{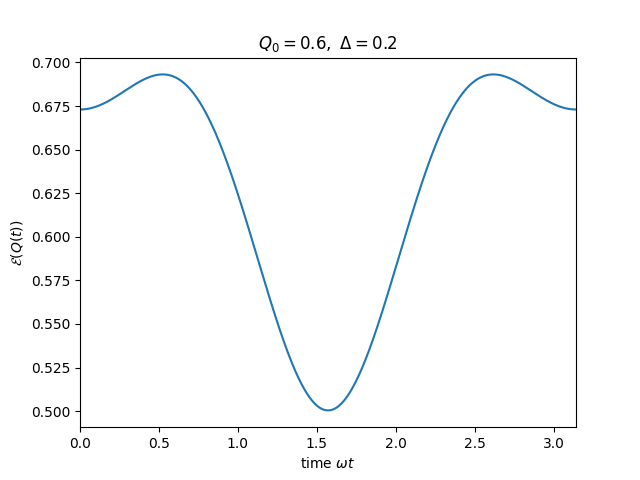}
	\end{subfigure}
	\caption{Parameters $Q_0=0.6$, $\Delta=\max_t\Delta(t)=0.2$. In this case $Q(t)$ starts at $Q_m=0.4$ when $t=0$ (upon a possible shift of the time axis) and moves to $Q_M=0.8$ at time $\omega t=\pi/2$ and back to $Q_m$ at time $\omega t=\pi$  (left panel). The value of the entropy $\mathcal E(Q(t))$ evolves through one single period (right panel). In each period, the entropy has two local minima (counting minima at the endpoints once).}
	\label{fig:test1}
\end{figure}

\begin{figure}[H]
	\centering
	\begin{subfigure}{.5\textwidth}
		\centering
		\includegraphics[width=1.2\linewidth]{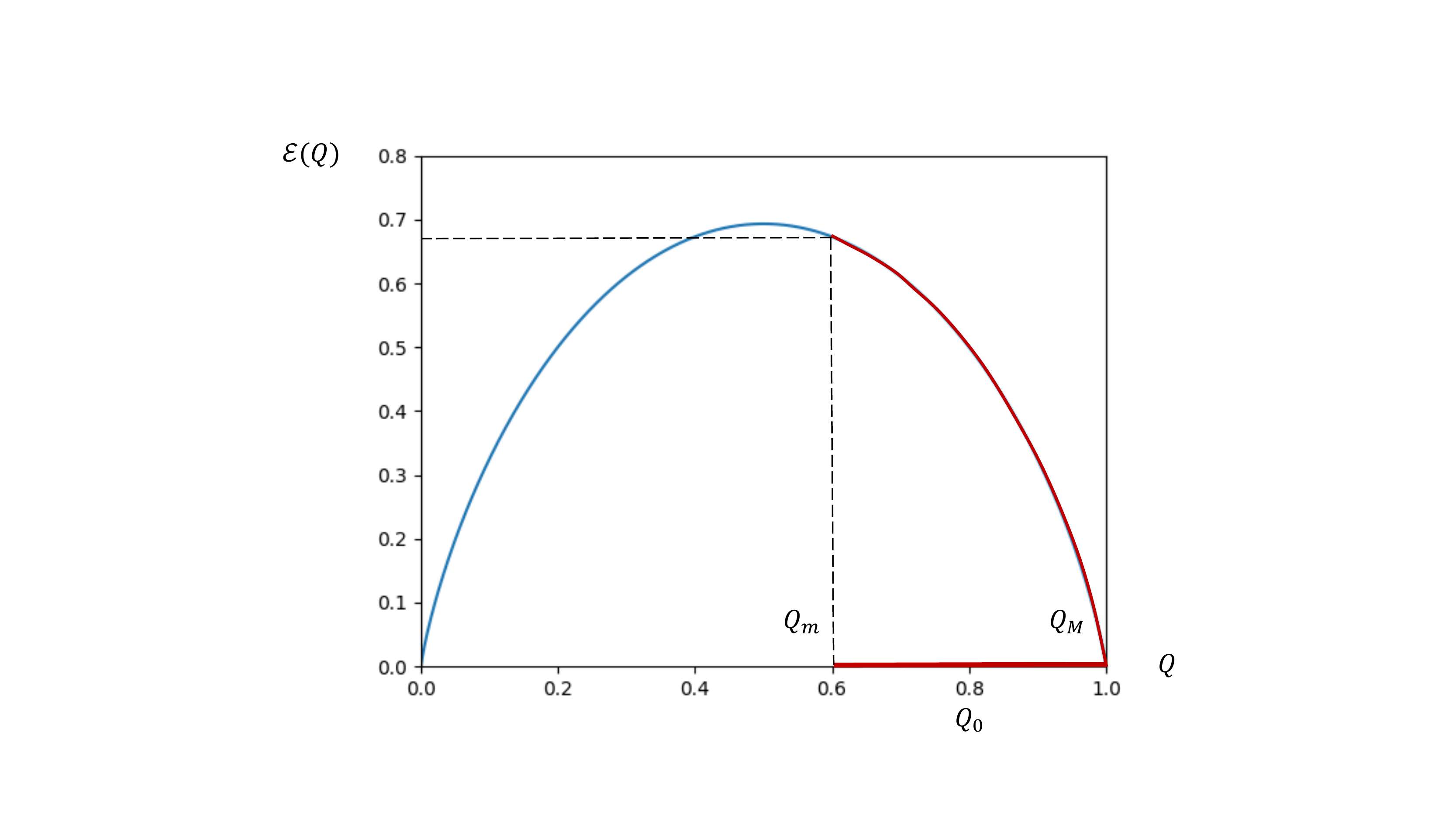}
  \vspace*{-1cm}
	\end{subfigure}%
	\begin{subfigure}{.5\textwidth}
		\centering
		\includegraphics[width=.75\linewidth]{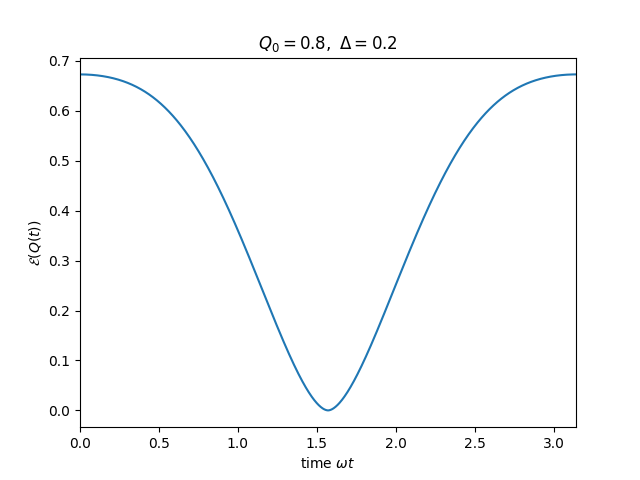}
	\end{subfigure}
	\caption{Parameters $Q_0=0.8$, $\Delta=\max_t\Delta(t)=0.2$. In this case $Q(t)$ starts at $Q_m=0.6$ when $t=0$ (after possibly shift the time axis) and moves to $Q_M=1.0$ at time $\omega t=\pi/2$ and back to $Q_m$ at time $\omega t=\pi$  (left panel). The value of the entropy $\mathcal E(Q(t))$ evolves through one single period (right panel). Since $0.5$ is not in the interval $(Q_m,Q_M)$, the graph of the entropy has only one local minimum in each period, instead of two when the interval contains the value $0.5$ for $Q$.}
	\label{fig:test2}
\end{figure}

On the other hand, if $Q_0\neq \frac12$, then the period of the entropy $\mathcal E(Q(t))$ is {\em not} doubled relative to that of $Q(t)$, as  Figures~\ref{fig:test1} and \ref{fig:test2} show.

We draw the following conclusions:

\begin{itemize}
\item[$\bullet$]  The period of the von Neumann entropy of the {\em non-Hermitian} system $\mathcal E(\bar\rho_H(t))$ is $\frac12\pi/\omega$, regardless of the initial condition (except for  the stationary state).

\item[$\bullet$] Regardless of the metric (parameter $x$), the period of the von Neumann entropy of the {\em Hermitian} system $\mathcal E(\bar\rho_{h_W}(t))$ is:
\begin{itemize}
	\item[$\circ$] $\pi/\omega$, provided ${\rm Re}(AB^*){\rm Re}(cd^*)\neq 0$ (generic case), 
	\item[$\circ$] $\frac12\pi/\omega$, provided ${\rm Re}(AB^*){\rm Re}(cd^*)=0$ (special case).
\end{itemize}
\end{itemize}
This means that for generic initial conditions (meaning ${\rm Re}(AB^*)\neq 0$) and generic choices of the unitary $W$ (meaning ${\rm Re}(cd^*)\neq 0$), {\bf the period of the von Neumann entropy of the Hermitian system is double that of the non-Hermitian system}. That is, the entropy of the non-Hermitian system oscillates faster. This is so even though the populations in both cases have the same frequency $\pi/\omega$. The change of the period is due to the shift of the average in the population induced by $W$, as given in \eqref{q0}.

\subsection{Numerical illustration of the period doubling}

We plot the populations and entropies for parameters in the regime
\begin{eqnarray}
	x=x_1=x_2&>&0\qquad \mbox{(metric $\eta$, \eqref{metricmatrix})}\label{m60.1}\\
	c,d &\ge& 0\qquad \mbox{(unitary $W$, \eqref{26})} \label{m61.2}\\
	A,B &\ge& 0\qquad \mbox{(initial state $|\psi(0)\rangle$,  \eqref{ic})} \label{m62.2}
\end{eqnarray}
According to \eqref{m61.2} and the unitarity of $W$, we have $d=\sqrt{1-c^2}$.  Moreover, $x^2A^2B^2=\alpha(1-\alpha)$, where
$\alpha = x\frac{a_2}{a_1}A^2\in[0,1]$. 
The population $q(t)$ of the Hermitian system reduced density matrix, given in \eqref{2-62}, then becomes
\begin{equation}
	\label{m61.1}
	q(t) = q_0 + \Delta \cos(2\omega t)
\end{equation}
with
\begin{eqnarray}q_0 &=& \tfrac12 +2 \sqrt{c^2(1-c^2)} \sqrt{\alpha(1-\alpha)} \label{m62.1},\\
	\Delta &=& \tfrac12 (1-2c^2)(1 - 2\alpha).
	\label{m63.1}
\end{eqnarray}
Here, $\alpha=x\frac{a_2}{a_1}A^2\in[0,1]$ and $c\in[0,1]$ can be chosen freely. The population $p(t)$ of the non-Hermitian system, in \eqref{2-49}, is simply the expression \eqref{m61.1} with $c=0$. 

Note that the change $\alpha\mapsto 1-\alpha$ leaves $q_0$ invariant and flips the sign of $\Delta$. It then suffices to plot graphs for $\alpha\in[0,1/2]$. For $\alpha=1/2$ we get $\Delta=0$, which gives a stationary state (for  all $c$). The same invariance of $q_0$ and sign flip of $\Delta$ is induced by $c^2\mapsto 1-c^2$.

\begin{figure}[H]
	\centering
\hspace*{-2.3cm}
\includegraphics[scale = 0.7]{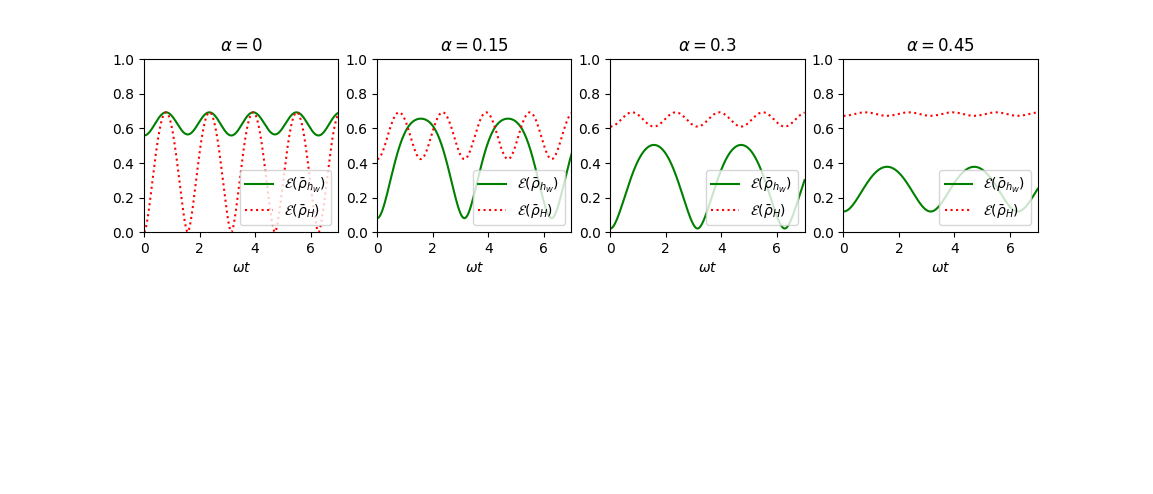}
\vspace*{-4cm}
	\caption{Comparing the entropies of $\bar\rho_{\hW}(t)$ and of $\bar\rho_H(t)$ for the values $c=0.5$ and $\alpha=0, 0.15, 0.3, 0.45$. The doubling of the period for $\alpha\neq 0$ is manifest. The oscillations decrease as $\alpha$ approaches $0.5$, which gives the stationary state.}
\label{fig:EhH}
\end{figure}

In Figure \ref{fig:EhH}, we compare the von Neumann entropies of the two density matrices $\bar\rho_H(t)$ and $\bar\rho_{\hW}(t)$, directly seeing the doubling of the period. 

\section{Entanglement of system and bath oscillators }
\label{sect:ent}

The total Hilbert space $\h=\h_\s\otimes\h_\b$ in \eqref{Htot} is bipartite, one part being the singled-out oscillator (system), the other being the remaining $N$ oscillators (bath). We say that a nonzero vector $|\psi\rangle\in\h$ is of {\em product form}, or {\em disentangled}, if $|\psi\rangle = |\psi_\s\rangle \otimes|\psi_\b\rangle$ for some $|\psi_\s\rangle\in\h_\s$ and some $|\psi_\b\rangle\in\h_\b$. We call a nonzero $|\psi\rangle\in\h$ {\em entangled} if it is not of product form. The notion of being entangled or not does not depend on the metric determining the inner product of $\h$. Nevertheless, the physical interpretation of entanglement in terms of independence of the subsystems $\s$ and $\b$ does depend on the metric.  The physical manifestation of disentangled states is the independence of the two subsystems $\s$ and $\b$. Namely, if the inner product of $\h$ is given by a metric of the form 
 $\eta=\Lambda_\s\otimes\Lambda_\b$ (a particular example being $\bbbone_\s\otimes\bbbone_\b$), then  {\em measurement outcomes} of observables on either of the subsystems are {\em independent random variables}. This follows because expectation values of observables $O_\s\otimes O_\b$ in a state $|\psi_\s\rangle\otimes|\psi_\b\rangle$ split into products,
$$
\langle \psi_\s\otimes\psi_\b| \eta (O_\s\otimes O_\b)\psi_\s\otimes\psi_\b\rangle = \langle\psi_\s| \Lambda_\s O_\s\psi_\s\rangle  \langle \psi_\b| \Lambda_\b O_\b\psi_\b\rangle.
$$
However, those random variables become dependent (correlated) if $\eta$ is not of product form, because then their average will not split into a product of a system term times a bath term.

 In the model defined in Section \ref{sec:model}, the metrics $\eta$ we consider are restrictions to the subspace $\h_1$ of product metrics $\Lambda_\s\otimes\Lambda_\b$ of $\h$ (c.f. \eqref{metricmatrix-1} and Section \ref{sect3.4}). The physical meaning of $\s\b$ entanglement in terms of subsystem independence does therefore not depend on the choice of $\eta$ within this class.  In other words, measurement outcomes of system and bath observables in $|\psi\rangle\in\h_1$ are independent or dependent, according to whether $|\psi\rangle\in\h_1$ is disentangled or not, regardless of the choice of $\eta$.  It is then sensible to investigate the SB entanglement in pure states belonging to the subspace $\h_1$ for all $\eta$.

Any vector $|\psi\rangle\in\h_1$ is of the form
\begin{equation}
	\label{2-73}
|\psi\rangle = A |e_\s\rangle +B |e_\b\rangle,\qquad A,B\in \mathbb C.
\end{equation}
In accordance with \eqref{m28.1}, we may write $|e_\s\rangle = |10\rangle = |1\rangle\otimes|0\rangle\in\h_\s\otimes\h_\b$ and $|e_\b\rangle=|01\rangle = |0\rangle\otimes|1\rangle\in\h_\s\otimes\h_\b$. Explicitly, $|0\rangle, |1\rangle\in\h_\s$ are the ground state and first excited state of the system oscillator, and $|0\rangle,|1\rangle\in\h_\b$ are the ground state of all the $N$ bath oscillators and the distributed excitation state $\frac{1}{\sqrt N}\sum_{n=1}^N|1_n\rangle$, respectively;  see \eqref{m28.1}.

We now show that 
\begin{equation}
\label{2-74}
|\psi\rangle \mbox{\ of the form \eqref{2-73} is disentangled}\quad\Longleftrightarrow\quad   AB =0.
\end{equation}
To see that the implication $\Rightarrow$ in \eqref{2-74} holds,  let $\rho_\s\equiv {\rm tr}_\b|\psi\rangle\langle\psi|$ (partial trace over $\b$). On the one hand, \eqref{2-73} gives $\rho_\s= |A|^2 |1\rangle\langle 1| +|B|^2 |0\rangle\langle 0|$. On the other hand, if  $|\psi\rangle$ is disentangled, then $\rho_\s$ must have rank one, since $\rho_\s=  {\rm tr}_\b ( |\psi_\s\rangle\langle\psi_\s|\otimes |\psi_\b\rangle\langle\psi_\b|)  = |\psi_\s\rangle\langle\psi_\s|\ \|\psi_\b\|^2$. This forces either $A=0$ or $B=0$. Conversely, to see the implication $\Leftarrow$ in \eqref{2-74}, we note that if either of  $A$ or $B$ vanish, then $|\psi\rangle$ is is proportional to $|e_\b\rangle$ or $|e_\s\rangle$, so $|\psi\rangle$ is disentangled.
\smallskip

\medskip

{\bf Entanglement in the non-Hermitian system.\ } An initial state $|\psi(0)\rangle =  A|e_\s\rangle +B|e_\b\rangle$ evolves into $|\psi(t)\rangle = e^{-i t H}|\psi(0)\rangle =A(t)|e_\s\rangle +B(t)|e_\b\rangle$, where the time dependent coefficients are given in \eqref{m39}. According to \eqref{2-74}, $|\psi(t)\rangle$ is disentangled exactly if $A(t)B(t)=0$. Let us first analyze the condition $A(t)=0$. This equality is equivalent to the two equations
\begin{equation}
\begin{aligned}
( {\rm Re}A )\cos(\omega t) + ({\rm Im}B) \frac{a_1}{a_2}\sin(\omega t) &= 0,\\
( {\rm Im}A )\cos(\omega t) - ({\rm Re}B) \frac{a_1}{a_2}\sin(\omega t) &= 0.
\end{aligned}
\label{2-81}
\end{equation}
The condition $B(t)=0$ is the same as \eqref{2-81} but with $A\leftrightarrow B$ swapped and $a_1\leftrightarrow a_2$ swapped. 

Suppose $|\psi(0)\rangle$ is disentangled, so $AB=0$. Then exactly one of $A$ or $B$ vanish and the equations \eqref{2-81} are  satisfied for $\omega t\in \pi \mathbb Z$ (if $A=0$) or for $\omega t = \frac\pi2 (2\mathbb Z+1)$ (if $B=0$). We conclude that $|\psi(t)\rangle$ is entangled except periodically at discrete moments in time where it is disentangled.

On the other hand, if $|\psi(0)\rangle$ is entangled, then both $A$ and $B$ do not vanish. If $A$ and $B$ are both real or both purely imaginary, then \eqref{2-81} is not satisfied for any $t$. For all other $A$ and $B$ \eqref{2-81} is satisfied for discrete, periodically repeating values of $t$.
\medskip  

We conclude:
\begin{itemize}
\item[(a)] If the initial state $|\psi(0)\rangle$ is entangled and both $A,B$ are either purely real or purely imaginary, then $|\psi(t)\rangle$ is entangled for all times $t$. 

\item[(b)] With the exception of case (a) and regardless of the entanglement in the initial state $|\psi(0)\rangle$, the state $|\psi(t)\rangle$ is entangled except at periodically repeating instants.
\end{itemize}

\medskip

\noindent {\bf Entanglement in the Hermitian systems.\ } The Hermitian system pure state vector is given by (see also \eqref{m19})
$$
|\phi(t)\rangle = S|\psi(t)\rangle = W\sqrt{\eta}|\psi(t)\rangle = \widetilde A(t) |e_\s\rangle +\widetilde B(t) |e_\b\rangle,
$$
a normalized vector in $\h$ (with the original inner product),  where
\begin{equation}
	\label{2-76}
\begin{pmatrix}
\widetilde A(t)\\
\widetilde B(t)
\end{pmatrix}
= T \begin{pmatrix}
	 A(t)\\
	 B(t)
\end{pmatrix},\qquad 
T=\sqrt{x\frac{a_2}{a_1 }} \begin{pmatrix}
a\ & ba_1/a_2\\
c   & d a_1/a_2
\end{pmatrix},\qquad \det T = x\det W\neq 0
\end{equation}
satisfy $|\widetilde A(t)|^2+ |\widetilde B(t)|^2=1$; c.f. \eqref{m49} and \eqref{26}. It follows from \eqref{2-74} that $|\phi(t)\rangle$ is entangled if and only if $\widetilde A(t)\widetilde B(t)=0$. An analysis of the latter equality along the lines of that carried out after \eqref{2-81} shows that $|\phi(t)\rangle$ is entangled except at isolated, periodically reoccurring instants in time,  just like the state of the non-Hermitian system.
\medskip

\bigskip

\noindent {\bf Effect of choice of $W$ on entanglement.\  } Given a state $|\psi\rangle=A|e_\s\rangle +B|e_\b\rangle$ of the non-Hermitian system, the associated Hermitian system state is $|\phi\rangle = W\sqrt\eta|\psi\rangle$. For the choice $W=\bbbone$ we have 
\begin{equation}
	\label{2-77}
|\phi\rangle = \sqrt\eta|\psi\rangle =\sqrt{xa_2/a_1} A|e_\s\rangle + \sqrt{x a_1/a_2}
B|e_\b\rangle.
\end{equation}
Hence for $W=\bbbone$, $|\phi\rangle$ is entangled if and only if $|\psi\rangle$ is entangled (recall \eqref{2-74}). The metric $\eta$ does not alter the property of being entangled. Choosing a different $W$ to build $|\phi\rangle$ from $|\psi\rangle$, however, changes this. It is not hard to see that the unitaries $W$ that map every product state (that is $|e_\s\rangle$ and $|e_\b\rangle$) into another product state are exactly the diagonal and the off-diagonal $W$. Furthermore, given an entangled $\sqrt\eta|\psi\rangle$ as in \eqref{2-77}, one can always find unitaries $W$ such that $W\sqrt\eta |\psi\rangle$ is not entangled. Those $W$ are precisely the ones with $|a|=|d|=\sqrt{a_1/(xa_2)}|B|$ and $|b|=|c|=\sqrt{xa_2/a_1}|A|$.\footnote{This follows from the characterization \eqref{2-74} of product states and the normalization $\|\sqrt\eta |\psi\rangle \|_\h=1$.} 
\medskip

We now examine the effect of $W$ on the {\em time-averaged density matrix}
\begin{equation}
\langle\rho\rangle = \frac\omega\pi \int_0^{\pi/\omega} |\phi(t)\rangle\langle \phi(t)|  \,dt
\end{equation}
where we integrate $|\phi(t)\rangle\langle \phi(t)|=\rho_{h_W}(t)$ over one period, see \eqref{rhwmatrix}. A direct calculation yields
\begin{equation}
	\label{2-80}
	\langle\rho\rangle =
\begin{pmatrix}
q_0 & z\\
 z^*& 1-q_0
\end{pmatrix},\quad 
 z=(bc^*+ad^*)x {\rm Re}(AB^*),
\end{equation}
with $q_0= \tfrac12 +2x{\rm Re}(AB^*){\rm Re}(cd^*)$, cf. \eqref{q0}. The density matrix \eqref{2-80} is written in the basis $\{|e_\s\rangle \equiv|10\rangle, |e_\b\rangle\equiv|01\rangle\}$ of $\h_1$, using the same notation as after \eqref{2-73}. We view $\h_1$ as a subspace of the four dimensional space of two qubits, spanned by the vectors $\{|00\rangle, |01\rangle, |10\rangle, |11\rangle\}$. In this basis, the density matrix \eqref{2-80} takes the form
\begin{equation}
\langle\rho\rangle =
\begin{pmatrix}
0 & 0 & 0 & 0\\
0 & 1-q_0 & z^* & 0\\
0 & z & q_0 & 0\\
0 & 0 & 0 & 0
\end{pmatrix}.
\end{equation}
We calculate the {\em concurrence} \cite{Wooters, Horodecki} of $\langle\rho\rangle$ to be\footnote{In the present case, the square of the concurrence is the difference between the two non-zero eigenvalues of the squared matrix $\langle\rho\rangle^2$.}
\begin{equation}
\mathcal C\big(\langle\rho\rangle\big) = 2x |{\rm Re}(AB^*)|. 
\end{equation}
The concurrence of any two qubit density matrix is bounded below by $0$ (separable state) and above by $1$ (maximally entangled state). Both inequalities in
$$
2x|{\rm Re}(AB^*)|\le 2x|A| |B| \le x\Big( \frac{a_2}{a_1}|A|^2 +\frac{a_1}{a_2}|B|^2\Big)=1
$$
(the last equality is the normalization \eqref{m38}, with $x=x_1=x_2>0$) are saturated exactly if $AB^*\in\mathbb R$ and $|A|=\frac{a_1}{a_2}|B|$. 

We conclude that the concurrence of $\langle\rho\rangle$  is the same for all choices $W$, so it only depends on the initial state. The state $\langle\rho\rangle$ is separable if and only if ${\rm Re}(AB^*)=0$, and is maximally entangled if and only if ${\rm Im}(AB^*)=0$  and $|A|=\frac{a_1}{a_2}|B|$.

\section{Conclusion}

The Dyson map assigns to a given quasi-Hermitian quantum system an associated Hermitian system in a non-unique way. We quantify the non-uniqueness by means of a metric operator $\eta$ and a unitary map $W$. The physical properties of the Hermitian systems depend on the choice of $W$, and it is not obvious how to capture the dynamics of the original quasi-Hermitian system in its Hermitian counterparts -- unless there happens to be some universality throughout the Hermitian family. We describe an aspect of universality for a quasi-Hermitian open system consisting of a single oscillator coupled to a bath of $N$ oscillators. We show that there is a unique metric operator for which the reduced state of the system (single oscillator) is a well-defined density matrix. Using this metric, we construct all Hermitian systems obtained from the quasi-Hermitian one by varying $W$. We find that the entropy of the single oscillator in the Hermitian system evolves periodically in time with exactly {\em double the period} of the corresponding entropy of the quasi-Hermitian system, independently of $W$. We further show that averaged over one time period, the entanglement between the oscillator and the bath is independent of the choice of $W$.

\appendix

\section{Appendix}

\subsection{Diagonal form of $\eta$ is equivalent to product form}
\label{sect3.4}

We have seen above in \eqref{metricmatrix} that $\eta$ must be diagonal in order for the populations of $\bar\rho_H(t)$  to be non-negative and that conversely if $\eta$ is diagonal, then the populations of $\bar\rho_H(t)$ are positive. As it turns out, $\eta$ being diagonal is also equivalent to $\eta$ being of product form. More precisely, the following two statements are equivalent:
\begin{itemize}
	\item[(1)] $\eta$ is diagonal in the basis $|e_\s\rangle,|e_\b\rangle$ of $\h_1$.
	\item[(2)] There are metrics $\Lambda_\s$ and $\Lambda_B$ on $\h_\s$ and $\h_\b$, respectively, such that $\Lambda_\s\otimes\Lambda_\b$ leaves $\h_1$ invariant and $\eta = \Lambda_\s\otimes\Lambda_\b \!\upharpoonright_{\h_1}$ is the restriction of this product to $\h_1$. 
\end{itemize}
Given $\eta$, the $\Lambda_\s$ and $\Lambda_\b$ are not unique. 

\bigskip

{\em Proof of (1) $\Leftrightarrow$ (2).}

Consider a bipartite Hilbert space $\h=\h_\s \otimes\h_\b$ with an orthonormal basis $|v_{ij}\rangle = |e_i\rangle\otimes |f_j\rangle$, the $|e_i\rangle$ and $|f_j\rangle$ being orthonormal bases of $\h_\s$ and $\h_\b$, respectively. Let $\h_1$ be the two-dimensional subspace $\h_1={\rm span}\{|v_{11}\rangle, |v_{22}\rangle\}$. In this setup, $|v_{11}\rangle$ is identified with $|e_\s\rangle$ and $|v_{22}\rangle$ with $|e_\b\rangle$. Let $\eta>0$ be a strictly positive operator on $\h_1$. 
\smallskip

We first show (1) $\Rightarrow$ (2). Assume that $\eta$ is diagonal, that is, $\eta = a|v_{11}\rangle\langle v_{11}| +b|v_{22}\rangle\langle v_{22}|$, where $a,b>0$. Set
\begin{eqnarray*}
	\Lambda_\s &=& \alpha_1|e_1\rangle\langle e_1| +\alpha_2|e_2\rangle\langle e_2| +\Lambda_\s^\perp\\
	\Lambda_\b &=& \beta_1|f_1\rangle\langle f_1| +\beta_2|f_2\rangle\langle f_2| +\Lambda_\b^\perp,
\end{eqnarray*}
where $\alpha_1,\alpha_2,\beta_1,\beta_2>0$ satisfy $\alpha_1\beta_1=a$, $\alpha_2\beta_2=b$ and $\Lambda_\s^\perp$, $\Lambda_\b^\perp$ are arbitrary positive operators on the orthogonal complements of ${\rm span}\{|e_1\rangle,|e_2\rangle\}$ and ${\rm span}\{|f_1\rangle,|f_2\rangle\}$ in $\h_\s$ and $\h_\b$, respectively. Then  $\Lambda_\s\otimes\Lambda_\b$ is a metric on $\h$ which leaves $\h_1$ invariant and satisfies $(\Lambda_\s\otimes\Lambda_\b)|v_{jj}\rangle = \eta|v_{jj}\rangle$ for $j=1,2$.

Next we prove that (2) $\Rightarrow$ (1). The orthogonal projection onto $\h_1$ is given by
$$
\pi = |v_{11}\rangle\langle v_{11}| +|v_{22}\rangle \langle v_{22}| = p_{11}\otimes q_{11} +p_{22}\otimes q_{22},
$$
where $p_{ij}=|e_i\rangle\langle e_j|$ and $q_{ij}=|f_i\rangle\langle f_j|$. 
Since $\Lambda_\s\otimes\Lambda_\b$ leaves $\h_1$ invariant, we have $(\Lambda_\s\otimes\Lambda_\b)\pi=\pi(\Lambda_\s\otimes\Lambda_\b)\pi$. Now
\begin{equation}
	\label{70}
	(\Lambda_\s\otimes\Lambda_\b)\pi = \Lambda_\s p_{11}\otimes \Lambda_\b q_{11} + \Lambda_\s p_{22}\otimes\Lambda_\b q_{22}
\end{equation}
and, with $[\Lambda_\s]_{ij}=\langle e_i,\Lambda_\s e_j\rangle$ and similarly for $\Lambda_\b$,
\begin{eqnarray}
	\pi( \Lambda_\s\otimes\Lambda_\b)\pi &=& [\Lambda_\s]_{11} [\Lambda_\b]_{11} \ p_{11}\otimes q_{11} + [\Lambda_\s]_{12} [\Lambda_\b]_{12}\  p_{12}\otimes q_{12}\nonumber\\
	&=&[\Lambda_\s]_{21} [\Lambda_\b]_{21} \ p_{21}\otimes q_{21} + [\Lambda_\s]_{22} [\Lambda_\b]_{22}\  p_{22}\otimes q_{22}.
	\label{71}
\end{eqnarray}
Taking the partial trace over $\b$ in \eqref{70} and \eqref{71} and equating the two results gives
\begin{equation}
	[\Lambda_\b]_{11} \Lambda_\s p_{11} + [\Lambda_\b]_{22} \Lambda_\s p_{22}
	=
	[\Lambda_\s]_{11} [\Lambda_\b]_{11}\  p_{11} + [\Lambda_\s]_{22} [\Lambda_\b]_{22} \ p_{22}. 
	\label{72}
\end{equation}
Since $[\Lambda_\b]_{11}, [\Lambda_\b]_{22}>0$ we get from \eqref{72} that $\Lambda_\s p_{jj}=[\Lambda_\s]_{jj}p_{jj}$ for $j=1,2$, so $\Lambda_\s |e_j\rangle = [\Lambda_\s]_{jj}|e_j\rangle$. Hence the restriction of $\Lambda_\s$ to ${\rm span}\{|e_1\rangle,|e_2\rangle\}$ is diagonal, $\Lambda_\s = [\Lambda_1]_{11} p_{11} + [\Lambda_\s]_{22} p_{22} +\Lambda^\perp_\s$, where $\Lambda_\s^\perp$ is the block acting on the orthogonal complement of that span. By taking the partial trace over $\s$ in \eqref{70} and \eqref{71} and proceeding analogously, we see that $\Lambda_\b=[\Lambda_\b]_{11} q_{11}+[\Lambda_\b]_{22}q_{22}+\Lambda_\b^\perp$. It follows that  $(\Lambda_\s\otimes\Lambda_\b) |v_{jj}\rangle = [\Lambda_\s]_{jj}[\Lambda_\b]_{jj}\,  |v_{jj}\rangle$ for $j=1,2$, 
so
$$
\eta = [\Lambda_\s]_{11}[\Lambda_\b]_{11}\,  |v_{11}\rangle\langle v_{11}| + 
[\Lambda_\s]_{22}[\Lambda_\b]_{22}\,  |v_{22}\rangle\langle v_{22}|
$$
is diagonal.

\subsection{Conditions for $\bar\rho_H(t)=\bar\rho_{\hW}(t)$ and $\mathcal E(\bar\rho_H(t))=\mathcal E(\bar\rho_{\hW}(t))$}
\label{AppB}

In this section we assume the metric $\eta$ is of the form \eqref{11} with $x_1=x_2=x>0$. Recall the formulas \eqref{m45} and \eqref{m56} for the quasi-Hermitian and Hermitian density matrices.
\medskip

First we ask when the two reduced density matrices coincide. We  show that the following statements are equivalent:
\begin{itemize}
\item[1.] $\bar\rho_H(t)=\bar\rho_{\hW}(t)$ for all $t$ in an open interval $I\subset\rx$ and all $A,B\in\mathbb C$;
\item[2.]  $\bar\rho_H(t)=\bar\rho_{\hW}(t)$ for all $t\in\rx$ and all $A,B\in\mathbb C$;
\item[3.] There are two real phases $\Phi_1$, $\Phi_2$, such that
	$$	W=
	\begin{pmatrix}
	e^{i\Phi_1} & 0 \\
	0 & e^{i\Phi_1}
	\end{pmatrix}.
	$$ 
\end{itemize}

1. $\Rightarrow$ 3. : Assume that $\bar\rho_H(t)=\bar\rho_{\hW}(t)$ for $t\in I$. Then $p(t)=q(t)$ for $t\in I$, where these quantities are given in \eqref{2-49} and \eqref{2-62}, respectively. Their equality is equivalent with
\begin{equation}
	\label{b1}
	\xi_1 +\xi_2\cos(2\omega t) +\xi_3 \sin(2\omega t) =0 \mbox{\quad for all $t\in I$},
\end{equation}
with $\xi_1=2x{\rm Re}(AB^*)\, {\rm Re} (c d^*)$, $\xi_2 = -|c|^2(1-2\alpha) -2x{\rm Im}(AB^*){\rm Im}(cd^*)$ and $\xi_3= -(1-2\alpha){\rm Im}(cd^*) +2x(1-|c|^2){\rm Im}(AB^*)$. As the constant function and the sine and cosine are three independent functions, and since \eqref{b1} holds for all $t$ in an interval, we conclude that  $\xi_1=\xi_2=\xi_3=0$. Since $\xi_1=0$ for all $A,B$ and $x>0$, we have ${\rm Re}(cd^*)=0$, that is, ${\rm Im}(cd^*)=cd^*$. The coefficients $A,B$ and $\alpha$ are related by \eqref{m38} and \eqref{alpha}, resulting in $A=\sqrt{\alpha}\sqrt{\frac{a_1}{xa_2}} e^{if_1}$ and $B=\sqrt{1-\alpha}\sqrt{\frac{a_2}{xa_1}}e^{if_2}$, where $f_1, f_2\in\mathbb R$ are phases. This gives ${\rm Im}(AB^*) = \sqrt{\alpha(1-\alpha)}\frac1x\, {\rm Im}\, e^{i(f_1-f_2)}$. Then $\xi_2=0$  for all $A,B$ implies that
$$
|c|^2(1-2\alpha) = -2\sqrt{\alpha(1-\alpha)} cd^* \, {\rm Im} \, e^{i(f_1-f_2)} \quad\mbox{for all $f_2,f_2\in\mathbb R$,  $\alpha\in[0,1]$.} 
$$
This forces $|c|^2(1-2\alpha)=0=\alpha(1-\alpha)cd^*$ for all $\alpha\in[0,1]$. Hence $c=0$. Then due to \eqref{26}, $|d|=1$ and $b d^*=0$, so $b=0$, and statement 3. holds.

3. $\Rightarrow$ 2. :  Suppose $c=0$. Then $|d|=1$ and from \eqref{m57} we have $q(t)=x\frac{a_1}{a_2}|B(t)|^2$, which equals the population of $|0_\s\rangle$ in $\bar\rho_H(t)$, see \eqref{19}. Therefore 2. holds.

2. $\Rightarrow$ 1. : Obvious.

This completes the proof of the equivalence of the three statements 1.-3.
\bigskip

Next we ask when the entropies of the two density matrices coincide. We show that the following statements 4.-6. are equivalent:
\begin{itemize}
	\item[4.] $\mathcal E(\bar\rho_H(t))=\mathcal E(\bar\rho_{\hW}(t))$  for all $t$ in an open interval $I\subset\rx$ and all $A,B\in\mathbb C$;
	\item[5.]  $\mathcal E(\bar\rho_H(t))=\mathcal E(\bar\rho_{\hW}(t))$  for all $t\in\rx$ and all $A,B\in\mathbb C$;
	\item[6.] There are two real phases $\Phi_1$, $\Phi_2$ such that $W$ is of either of the two forms
	$$
	W=
	\begin{pmatrix}
	e^{i\Phi_1} & 0 \\
	0 & e^{i\Phi_1}
	\end{pmatrix}\qquad
	\text{ or }\qquad   W=
	\begin{pmatrix}
	0 & e^{i\Phi_1}   \\
	e^{i\Phi_2} & 0  
	\end{pmatrix}.
	$$
\end{itemize}

4. $\Rightarrow$ 6. :  Start by looking at the function $\mathcal E(q)= -q\ln(q)-(1-q)\ln(1-q)$, for  $q\in[0,1]$. It is clear from the graph of $\mathcal E(q)$ (see the left panel of Figure~\ref{fig:test}) that $\mathcal E(q)=\mathcal E(q')$ exactly if either $q=q'$ or $q=1-q'$. Consequently, if  $\e(\bar\rho_{\hW}(t))=\e(\bar\rho_H(t))$ for all $t\in I$, then for each $t\in I$ {\em individually}, we have either $p(t)=q(t)$ or $p(t)=1-q(t)$. We now show that the same alternative must happen for {\em all} $t\in I$.

Suppose first that $p(t_0)\neq 1-q(t_0)$ for some $t_0\in I$. Then by the continuity of $p(t)$ and $q(t)$, we have $p(t)\neq 1-q(t)$ for all $t$ in an open interval $I_0\subset I$ around $t_0$, so we must have $p(t)=q(t)$ for $t\in I_0$. But this means that $\bar\rho_H(t)=\bar\rho_{\hW}(t)$ for all $t\in I_0$. Hence, as 2. and 3. are equivalent, $W$ is of the diagonal form as given in point 3. above. Similarly, if $p(t_0)\neq q(t_0)$ for some $t_0\in I$, we obtain $p(t) = 1-q(t)$ an an interval around $t_0$. Proceeding as in the proof of the implication 1. $\Rightarrow$ 3. above, this implies that $a=d=0$, so $W$ is of the off-diagonal form given in statement 6. above. 

6. $\Rightarrow$ 5. : If $W$ is of the diagonal form, then we already showed  that $p(t)=q(t)$ when we proved 3. $\Rightarrow$ 2. In the same way, if $W$ is off-diagonal, then one sees that $p(t)=1-q(t)$. In either case, $\mathcal E(\bar\rho_H(t))=\mathcal E(\bar\rho_{\hW}(t))$.

5. $\Rightarrow$ 4. : Obvious.

\bigskip

\subsection{About the Dyson map}
\label{sec:dysonmap}

The idea of mapping a non-Hermitian Hamiltonian to a Hermitian one was originally presented by Dyson in the context of the theory of magnetization \cite{Dyson56-1, Dyson56-2}. Let $\h$ be a Hilbert space with inner product $\langle\cdot|\cdot\rangle$ and let $H$ be an operator on $\h$ that is not necessarily Hermitian with respect to $\langle\cdot|\cdot\rangle$. Denote by $|\psi(t)\rangle = e^{-i t H}|\psi(0)\rangle$ the solution of the evolution equation
\begin{equation}
	\label{m7}
	i\partial_t|\psi(t)\rangle = H|\psi(t)\rangle.
\end{equation}
Next, let $S(t)$ be a differentiable family of operators on $\h$ such that $S(t)$ is invertible for each $t$, and set 
\begin{equation}
	|\varphi(t)\rangle = S(t)|\psi(t)\rangle.
\end{equation}
Passing from $|\psi(t)\rangle$ to $|\varphi(t)\rangle$ represents a (possibly time-dependent) change of variables. The evolution equation for $|\varphi(t)\rangle$ is
\begin{equation}
	\label{m9}
	i \partial_t |\varphi(t)\rangle = h(t)|\varphi(t)\rangle,
\end{equation}
with
\begin{equation}
	\label{m10}
	h(t) = S(t) H S(t)^{-1} + i\dot S(t)S(t)^{-1},
\end{equation}
the dot being the time derivative. Conversely, if $|\varphi(t)\rangle$ solves \eqref{m9} then $|\psi(t)\rangle$ solves \eqref{m7}. Equation \eqref{m10} is called the {\em time-dependent Dyson equation} \cite{FF19}. By means of $S(t)$, one may thus equivalently solve \eqref{m7} or \eqref{m9}. If $H$ is not Hermitian, one can look for $S(t)$ such that the resulting $h(t)$ is Hermitian, hence trading a non-Hermitian problem with constant Hamiltonian $H$ for a Hermitian problem with time-dependent Hamiltonian $h(t)$.  One readily sees that
\begin{equation} 
	\label{m11}
	h(t)^\dagger=h(t) \quad  \Longleftrightarrow \quad  i\partial_t \big(S^\dagger(t)S(t)\big) = H^\dagger \big( S^\dagger(t)S(t)\big) - \big(S^\dagger(t)S(t)\big) H.
\end{equation}
The operator $\eta(t)=S(t)^\dagger S(t)$ is automatically positive, so $\eta(t)$ is a family of metrics.  The  equation for $\eta(t)$, according to  \eqref{m11}, is 
\begin{equation}
	\label{m12}
	i\partial_t \eta(t) = H^\dagger \eta(t) - \eta(t)H.
\end{equation}
This is called the {\em quasi-Hermiticity relation} in \cite{FF17}; note that it simplifies to \eqref{m1} if $\eta$ does not depend on time. It is clear that \eqref{m12} has a unique solution for any initial condition $\eta(0)$, namely
\begin{equation}
	\label{m13}
	\eta(t) = e^{-i tH^\dagger} \eta(0) e^{i t H},
\end{equation}
and that $\eta(t)$ is positive for all times if and only if it is positive at some  $t_0$. 

A strategy to study the dynamics generated by a non-Hermitian $H$ is to find a transformation $S(t)$ such that $h(t)$, as given by \eqref{m10}, is Hermitian, and then analyze the dynamics of this Hermitian system using usual quantum theoretical methods. Finding $S(t)$ for a specific Hamiltonian $H$ is not easy, however. It often involves making a judicious ansatz containing parameters that must solve rather complicated differential equations, which are obtained by imposing the self-adjointness of $h(t)$. 
This can be done explicitly for some models \cite{CF2009,FF17,FF19,Mos2005, Mum2007, FT2020, FT2021}.


Given $H$, we seek \emph{all} possible $S(t)$, and the resulting Hermitian Hamiltonians $h(t)$, with the sole requirement that $\eta(t) = S(t)^\dagger S(t)$ is positive and satisfies the quasi-Hermiticity relation \eqref{m12}. The solution $\eta(t)$ is  uniquely determined by the initial condition $\eta(0)$, which we may choose to be any positive, invertible operator. The most general form of $S(t)$ is thus
\begin{equation}
	\label{m14}
	S(t)  = W(t) \sqrt{\eta(0)} e^{i t H},
\end{equation}
where $W(t)$ is any unitary family 
and $\sqrt{\eta(0)}$ denotes the unique positive  operator squaring to $\eta(0)$. The $h(t)$ associated to \eqref{m14} by \eqref{m10} is
\begin{equation}
	\label{m15}
	h(t) = i \dot W(t) W(t)^\dag.
\end{equation}
Note that we are entirely free to choose $W(t)$. For instance, given an arbitrary $A = A^\dagger$, the choice $W(t)=e^{- i t A}$ yields $h(t)=A$. This means any time-independent Hermitian $h$ can be obtained from a suitable choice of $S(t)$. A particularly simple choice is $S(t) = e^{itH}$, which results from undoing the dynamics $e^{-i t H}$ (going backwards in time) and has $h=0$.

More generally, suppose $A(t)$ is a continuous family of operators, and let $W(t)$ solve the differential equation
\begin{equation}
	\label{m16}
	i\dot W(t)=A(t)W(t).
\end{equation}
It is easily shown that if $A(t)=A(t)^\dagger$ for all $t$ and the initial condition $W(0)$ is unitary, then the solution $W(t)$ is unitary for all $t$. Choosing this $W(t)$, we find from \eqref{m15} the Hermitian Hamiltonian $h(t) = i \dot W(t) W(t)^\dag = i \dot W(t) W(t)^{-1} = A(t)$. This means any time-\emph{dependent} Hermitian $h(t)$ can also be obtained from a suitable choice of $S(t)$. 
\bigskip

{\bf Acknowledgements. } The authors are grateful to Andreas Fring for graciously explaining  the ideas and results of \cite{FF19}. All authors are supported by Discovery Grants from the Natural Sciences and Engineering Research Council of Canada (NSERC).

\end{document}